\documentstyle{mn}
\input{psfig}
\newcommand{\etal}{{et al.~}}

\newcommand{\kmsmpc}{\>{\rm km}\,{\rm s}^{-1}\,{\rm Mpc}^{-1}}
\newcommand{\kms}{\>{\rm km}\,{\rm s}^{-1}}

\newcommand{\Mpc}{\>{\rm Mpc}}

\newcommand{\Msun}{\>{\rm M_{\odot}}}
\newcommand{\Lsun}{\>{\rm L_{\odot}}}
\newcommand{\MLsun}{\>({\rm M}/{\rm L})_{\odot}}

\newcommand{\walpha}{\tilde{\alpha}}
\newcommand{\wLstar}{\tilde{L}^{*}}

\newcommand{\beq}{\begin{equation}}
\newcommand{\eeq}{\end{equation}}

\newcommand{\msunh}{\>h^{-1}\rm M_\odot}

\def\gtsima{$\; \buildrel > \over \sim \;$}
\def\ltsima{$\; \buildrel < \over \sim \;$}
\def\prosima{$\; \buildrel \propto \over \sim \;$}
\def\gsim{\lower.7ex\hbox{\gtsima}}
\def\lsim{\lower.7ex\hbox{\ltsima}}
\def\simgt{\lower.7ex\hbox{\gtsima}}
\def\simlt{\lower.7ex\hbox{\ltsima}}
\def\simpr{\lower.7ex\hbox{\prosima}}
\def\la{\lsim}
\def\ga{\gsim}
\def\lta{\la}
\def\gta{\ga}

\newcommand{\apj}{ApJ}

\newcommand{\aj}{AJ}
\newcommand{\mnras}{MNRAS}

\newcommand{\nat}{Nature}

\newdimen\hssize
\newdimen\hdsize
\begin{document}


\title[Probing Dark Matter Haloes with Satellite Kinematics]
      {Probing Dark Matter Haloes with Satellite Kinematics}
\author[van den Bosch, Norberg, Mo \& Yang]
       {Frank C. van den Bosch$^{1}$, Peder Norberg$^{1}$, 
        H.J. Mo$^{2}$, and Xiaohu Yang$^{2}$
        \thanks{E-mail: vdbosch@phys.ethz.ch}\\
        $^1$Department of Physics, Swiss Federal Institute of
         Technology, ETH H\"onggerberg, CH-8093, Zurich,
         Switzerland\\ 
        $^2$Department of Astronomy, University of Massachussets, 710
         North Pleasant Street, Amherst MA 01003-9305, USA}


\date{}

\pubyear{2004}

\maketitle

\label{firstpage}


\begin{abstract}
  Using detailed  mock galaxy redshift surveys we  investigate to what
  extent  the  kinematics  of  large  samples  of  satellite  galaxies
  extracted from  flux-limited surveys can  be used to  constrain halo
  masses.  Unlike previous studies,  which focussed only on satellites
  around  relatively isolated  host galaxies,  we try  to  recover the
  average velocity dispersion of  satellite galaxies in all haloes, as
  a  function of  the luminosity  of the  host galaxy.   We  show that
  previous  host-satellite selection  criteria yield  relatively large
  fractions  of interlopers  and  with a  velocity distribution  that,
  contrary to what has been assumed in the past, differs strongly from
  uniform.   We  show  that  with  an  iterative,  adaptive  selection
  criterion one can obtain large samples of hosts and satellites, with
  strongly  reduced  interloper  fractions,  that  allow  an  accurate
  measurement of $\sigma_{\rm sat}(L_{\rm  host})$ over two and a half
  orders  of magnitude  in host  luminosity.  We  use  the conditional
  luminosity  function  (CLF)  to  make  predictions,  and  show  that
  satellite  weighting,  which  occurs  naturally when  stacking  many
  host-satellite pairs to  increase signal-to-noise, introduces a bias
  towards  higher  $\sigma_{\rm sat}(L_{\rm  host})$  compared to  the
  true, host-averaged mean.  A further bias, in the same direction, is
  introduced  when  using  flux-limited,  rather  than  volume-limited
  surveys. We apply our adaptive selection criterion to the 2dFGRS and
  obtain a sample of 12569  satellite galaxies and 8132 host galaxies. 
  We  show that  the kinematics  of  these satellite  galaxies are  in
  excellent  agreement with the  predictions based  on the  CLF, after
  taking account of  the various biases.  We thus  conclude that there
  is  independent  dynamical  evidence  to support  the  mass-to-light
  ratios predicted by the CLF formalism.
\end{abstract}




\section{Introduction}
\label{sec:intro}

Ever since the ``discovery'' of dark matter astronomers have attempted
to obtain  accurate measurements  of the masses  of the  extended dark
matter haloes  in which  galaxies are thought  to reside.   A detailed
knowledge of  halo masses  around individual galaxies  holds important
clues  to  the  physics  of  galaxy formation,  and  is  an  essential
ingredient of any successful model that aims at linking the observable
Universe (i.e., galaxies) to  the bedrock of our theoretical framework
(i.e.  dark matter).

The main challenge  in measuring {\it total} halo masses  is to find a
suitable,  visible tracer  at  sufficiently large  radii  in the  halo
potential well.  Traditionally, starting  with the actual discovery of
evidence for dark matter by Zwicky (1933, 1937), astronomers have used
the kinematics of satellite  galaxies.  Since the number of detectable
satellites in individual systems is generally small, this technique is
basically limited to clusters  of galaxies (e.g., Carlberg \etal 1996;
Carlberg, Yee \& Ellingson 1997b) and the local group (e.g., Little \&
Tremaine 1987;  Lin, Jones \&  Klemola 1995; Evans \&  Wilkinson 2000;
Evans  \etal  2000).   However,  one   can  stack  the  data  on  many
host-satellite  pairs to  obtain {\it  statistical} estimates  of halo
masses.  Pioneering  efforts in this direction were  made by Erickson,
Gottesman \& Hunter (1987),  Zaritsky \etal (1993, 1997), and Zaritsky
\&  White (1994).  Although  these studies  were typically  limited to
samples  of less than  100 satellites,  they nevertheless  sufficed to
demonstrate  the  existence of  extended  massive  dark haloes  around
(spiral) galaxies.

More recently,  large, homogeneous galaxy  surveys, such as  the Sloan
Digital Sky  Survey (SDSS; York \etal  2000) and the  Two Degree Field
Galaxy Redshift Survey (2dFGRS;  Colless \etal 2001) have dramatically
improved both the  quantity and quality of data  on (nearby) galaxies,
thus   allowing   the  construction   of   much   larger  samples   of
host-satellite pairs (McKay \etal  2002; Prada \etal 2003; Brainerd \&
Specian  2003).   Yet,  each  of  these  studies  has  been  extremely
conservative in  their selection of ``isolated''  hosts and ``tracer''
satellites.  For  example, McKay \etal (2002), Prada  \etal (2003) and
Brainerd  \& Specian  (2003) used  samples with  1225, 2734,  and 2340
satellites, respectively.   For comparison, the SDSS  and 2dFGRS, from
which  these samples  were selected,  contain  well in  excess of  one
hundred thousand galaxies.  The  main reason for being so conservative
is to prevent too large  numbers of interlopers (i.e., satellites that
are  not bound  to their  host  galaxies, but  only appear  associated
because of  projection effects).  In addition,  the selection criteria
are  optimized  to  only  select  {\it isolated}  galaxies,  with  the
motivation that the dynamics of  binary systems, for example, are more
complicated.

In  this paper we  use a  different approach  and investigate  to what
extent the  kinematics of satellite  galaxies may be used  to estimate
the  mean mass-to-light ratio,  {\it averaged  over all  possible dark
  matter haloes}.   We make the simple ansatz  that satellite galaxies
are in  virial equilibrium  within their dark  matter potential  well. 
This is motivated by the finding that dark matter subhaloes, which are
likely  to be  associated with  satellite galaxies,  have  indeed been
found to  be in a  steady-state equilibrium (Diemand, Moore  \& Stadel
2004).  Although  we acknowledge  that not all  systems will  be fully
virialized, we  hypothesize that with  a sufficiently large  sample of
hosts  and   satellites  the  assumption  of   virial  equilibrium  is
sufficiently  accurate  to  describe  the  mean  properties.   We  use
detailed mock  galaxy redshift  surveys (hereafter MGRSs)  to optimize
the selection criteria  for host and satellite galaxies  and show that
an  iterative, adaptive  selection  criterion is  ideal  to limit  the
number of interlopers while still  yielding large numbers of hosts and
satellites.  Applying our selection criteria to the 2dFGRS yields 8132
hosts with  12569 satellites.   We show that  the kinematics  of these
satellite galaxies are in good agreement with predictions based on the
conditional  luminosity function  introduced by  Yang, Mo  \&  van den
Bosch (2003a) and van den Bosch, Yang \& Mo (2003a).

This  paper is  organized  as follows.   In Section~\ref{sec:mock}  we
first  describe,  in  detail,   the  construction  of  our  MGRSs.  In
Section~\ref{sec:method} we use these MGRSs to compare three different
selection criteria  for host  and satellite galaxies.   We investigate
the  impact of  interlopers  and non-central  hosts  on the  satellite
kinematics,  and  show how  stacking  data  from flux-limited  surveys
results  in a  systematic overestimate  of the  true  average velocity
dispersion of satellite galaxies.  We show  how the CLF can be used to
take these biases into  account.  In Section~\ref{sec:2dFres} we apply
our  selection criteria  to  the  2dFGRS and  compare  the results  to
analytical estimates  and to  our MGRS. We  summarize our  findings in
Section~\ref{sec:summ}.

Throughout   we   assume    a   flat   $\Lambda$CDM   cosmology   with
$\Omega_m=0.3$,  $\Omega_{\Lambda}=0.7$, $h=H_0/(100  \kmsmpc)  = 0.7$
and with  initial density fluctuations described  by a scale-invariant
power spectrum with  normalization $\sigma_8=0.9$.  
 
\section{Mock Galaxy Redshift Surveys}
\label{sec:mock}

What  is the  best  way to  select  host and  satellite galaxies  from
redshift  surveys such  as  the 2dFGRS  and  the SDSS?   How does  the
flux-limited nature  of these surveys  impact on the results?   How do
interlopers bias  the mass  estimates? In order  to address  these and
other questions  we use detailed mock galaxy  redshift surveys.  These
have the  advantage that (i)  we know exactly the  input mass-to-light
ratios  that  we   aim  to  recover,  (ii)  we   can  mimic  realistic
host/selection criteria and investigate the impact of interlopers.  In
addition,  MGRSs  allow a  detailed  investigation  of  the effect  of
Malmquist  bias,  various   survey  incompleteness  effects,  boundary
effects due to the survey geometry, etc.

To  construct MGRSs two  ingredients are  required; a  distribution of
dark  matter haloes  and a  description of  how galaxies  of different
luminosity occupy  haloes of  different mass.  For  the former  we use
large numerical simulations (see Section~\ref{sec:simulations} below),
and for  the latter the conditional luminosity  function $\Phi(L \vert
M) {\rm d}L$.  The conditional luminosity function (hereafter CLF) was
introduced by Yang \etal (2003a) and  van den Bosch \etal (2003a) as a
statistical tool  to link  galaxies to their  dark matter  haloes, and
describes the average  number of galaxies with luminosity  $L \pm {\rm
  d}L/2$ that reside in a halo  of mass $M$. As shown in these papers,
the  CLF is  well constrained  by  the 2dFGRS  luminosity function  of
Madgwick \etal  (2002) and  the correlation lengths  as a  function of
luminosity obtained  by Norberg \etal  (2002a). Details about  the CLF
used in this paper can be found in Appendix~A.

\subsection{Numerical Simulations}
\label{sec:simulations}

The distribution of dark matter haloes is obtained from a set of large
$N$-body   simulations   (dark  matter   only)   for  a   $\Lambda$CDM
`concordance' cosmology with $\Omega_m = 0.3$, $\Omega_{\Lambda}=0.7$,
$h=0.7$  and $\sigma_8=0.9$.   The  set  consists of  a  total of  six
simulations with  $N=512^3$ particles each,  and is described  in more
detail  in Jing  (2002)  and  Jing \&  Suto  (2002).  All  simulations
consider boxes with periodic boundary conditions; in two cases $L_{\rm
  box}=100  h^{-1} \Mpc$  while the  other four  simulations  all have
$L_{\rm box}=300  h^{-1} \Mpc$.   Different simulations with  the same
box  size are  completely  independent realizations  and  are used  to
estimate uncertainties  due to  cosmic variance.  The  particle masses
are $6.2  \times 10^8 \msunh$  and $1.7\times 10^{10} \msunh$  for the
small  and large box  simulations, respectively.   In what  follows we
refer to  simulations with $L_{\rm  box}=100 h^{-1} \Mpc$  and $L_{\rm
  box}=300  h^{-1}  \Mpc$  as  $L_{100}$  and  $L_{300}$  simulations,
respectively.

Dark    matter   haloes    are   identified    using    the   standard
friends-of-friends algorithm (Davis \etal  1985) with a linking length
of  $0.2$   times  the  mean  inter-particle   separation.   For  each
individual  simulation we construct  a catalogue  of haloes  with $10$
particles or  more, for which we  store the mass, the  position of the
most  bound  particle,  and  the  halo's mean  velocity  and  velocity
dispersion. Haloes  that are unbound  are removed from the  sample. In
Yang  \etal  (2003b)  we  have  shown that  the  resulting  halo  mass
functions  are in excellent  agreement with  the analytical  halo mass
function  given by Sheth,  Mo \&  Tormen (2001a)  and Sheth  \& Tormen
(2002).

\subsection{Halo Occupation Numbers}
\label{sec:hon}

Because of the  mass resolution of the simulations  and because of the
completeness limit of the 2dFGRS  we adopt a minimum galaxy luminosity
of $L_{\rm min} = 3  \times 10^{7} h^{-2} \Lsun$ throughout.  The {\it
  mean} number of galaxies with $L \geq L_{\rm min}$ that resides in a
halo of mass $M$ follows from the CLF according to:
\begin{equation}
\label{meanN}
\langle N \rangle_M = \int_{L_{\rm min}}^{\infty} \Phi(L \vert M) \,
{\rm d}L\,.
\end{equation}
In  order  to Monte-Carlo  sample  occupation  numbers for  individual
haloes one  requires the full probability distribution  $P(N \vert M)$
(with $N$ an integer) of  which $\langle N \rangle_M$ gives the mean,
i.e.,
\begin{equation}
\label{meanNint}
\langle  N \rangle_M = \sum_{N=0}^{\infty} N \, P(N \vert M)
\end{equation}

We use  the results of Kravtsov  \etal (2003), who has  shown that the
number  of {\it subhaloes}  follows a  Poisson distribution.   In what
follows  we   differentiate  between  satellite   galaxies,  which  we
associate  with these  dark  matter subhaloes,  and central  galaxies,
which we associate with the host halo (cf. Vale \& Ostriker 2004). The
total number  of galaxies per  halo is the  sum of $N_{\rm  cen}$, the
number of  central galaxies which is  either one or  zero, and $N_{\rm
  sat}$, the (unlimited) number of satellite galaxies.  We assume that
$N_{\rm  sat}$  follows a  Poissonian  distribution  and require  that
$N_{\rm  sat}=0$  whenever   $N_{\rm  cen}=0$.   The  halo  occupation
distribution  is thus specified  as follows:  if $\langle  N \rangle_M
\leq 1$ then $N_{\rm sat} =  0$ and $N_{\rm cen}$ is either zero (with
probability $P = 1 - \langle N \rangle_M$) or one (with probability $P
= \langle  N \rangle_M$).  If $\langle  N \rangle_M >  1$ then $N_{\rm
  cen}=1$ and $N_{\rm sat}$ follows the Poisson distribution
\begin{equation}
\label{poisson}
P(N_{\rm sat} \vert M) = {\rm e}^{-\mu} 
{\mu^{N_{\rm sat}} \over N_{\rm sat}!}\,,
\end{equation}
with $\mu = \langle N_{\rm sat}  \rangle_M = \langle N \rangle_M - 1$. 
As discussed in Kravtsov \etal  (2003) the resulting $P(N \vert M)$ is
significantly  sub-Poissonian   for  haloes  with   small  $\langle  N
\rangle_M$  (i.e.,  low  mass  haloes), but  approaches  a  Poissonian
distribution for  haloes with large $\langle N  \rangle_M$.  Such $P(N
\vert   M)$  is   supported   by  both   semi-analytical  models   and
hydrodynamical simulations of structure formation (Seljak 2000; Benson
\etal 2000; Scoccimarro \etal 2001;  Berlind \etal 2003), and has been
shown  to  yield  correlation   functions  in  better  agreement  with
observations than, for  example, for a pure Poissonian  $P(N \vert M)$
(Benson \etal 2000; Berlind \& Weinberg 2002; Yang \etal 2003a).

\subsection{Assigning galaxies their luminosity and type}
\label{sec:luminosity}

Since the  CLF only  gives the {\it  average} number of  galaxies with
luminosities in  the range $L \pm {\rm  d}L/2$ in a halo  of mass $M$,
there are many different ways  in which one can assign luminosities to
the $N_i$  galaxies of halo  $i$ and yet  be consistent with the  CLF. 
The simplest approach would be to simply draw $N_i$ luminosities (with
$L > L_{\rm min}$) from $\Phi(L \vert M)$. We refer to this luminosity
sampling as `random'.  Alternatively, one could use a more constrained
approach,  and, for  instance,  always demand  that  the $j^{\rm  th}$
brightest  galaxy has  a luminosity  in the  range  $[L_{j},L_{j-1}]$. 
Here $L_j$  is defined such  that a halo  has on average  $j$ galaxies
with $L > L_j$, i.e.,
\begin{equation}
\label{Lj}
\int_{L_j}^\infty \Phi(L\vert M) dL = j\,.
\end{equation}
We refer to this luminosity sampling as `constrained'. 

We  follow Yang  \etal (2003b)  and  adopt an  intermediate approach.  
Throughout we assume  that the central galaxy is  the brightest galaxy
in  each halo,  and we  draw  its luminosity,  $L_c$, constrained.  It
therefore has an expectation value of
\begin{equation}
\label{Lcentral}
\langle L_c \rangle_M = \int_{L_1}^{\infty} \Phi(L \vert M) \, L \, {\rm d}L 
\end{equation}
The remaining $N_i-1$ satellite  galaxies are assigned luminosities in
the  range  $L_{\rm  min}  <  L  <  L_1$  drawn  at  random  from  the
distribution function $\Phi(L\vert M)$.

\subsection{Assigning galaxies their phase-space coordinates}
\label{sec:position}

Next the  mock galaxies  need to be  assigned a position  and velocity
within their  halo. We assume  that each dark  matter halo has  an NFW
density distribution (Navarro, Frenk \& White 1997) with virial radius
$r_{\rm vir}$, characteristic scale radius $r_s$, and concentration $c
=   r_{\rm  vir}/r_s$.    Throughout  this   paper  we   compute  halo
concentrations as a function of  halo mass using the relation given by
Eke,  Navarro  \&  Steinmetz   (2001),  properly  accounting  for  our
definition  of halo mass\footnote{Throughout  this paper  halo masses,
  denoted by $M$, are defined  as the mass inside the radius $R_{180}$
  inside of which the average overdensity is $180$.}.  The ``central''
(brightest) galaxy in  each halo is assumed to be  located at the halo
centre,  which  we associate  with  the  position  of the  most  bound
particle.  Satellite  galaxies are assumed  to follow a  radial number
density distribution given by
\begin{equation}
\label{nsatr}
n_{\rm sat}(r) \propto \left( {r \over {\cal R} r_s} \right)^{-\alpha}
\left( 1 + {r \over {\cal R} r_s} \right)^{\alpha-3}
\end{equation}
(limited to  $r \leq  r_{\rm vir}$) with  $\alpha$ and ${\cal  R}$ two
free parameters. Unless specifically stated otherwise we adopt $\alpha
= {\cal R} = 1$ for which the number density distribution of satellite
galaxies  exactly  follows  the  dark  matter  mass  distribution. 

Finally, peculiar velocities are  assigned as follows.  We assume that
the ``central''  galaxy is located at  rest with respect  to its halo,
and  set  its peculiar  velocity  equal to  the  mean  halo velocity.  
Satellite  galaxies are assumed  to be  in a  steady-state equilibrium
within the  dark matter potential well with  an isotropic distribution
of velocities  with respect to the  halo centre.  As  shown by Diemand
\etal (2004) this  is a good approximation for  dark matter subhaloes,
and we assume it also  applies to satellite galaxies.  One dimensional
velocities are drawn from a Gaussian
\begin{equation}
\label{gasv}
f(v_j)={1\over \sqrt{2\pi}\sigma_{\rm sat}(r)}
\exp \left( -{v_j^2 \over 2\sigma_{\rm sat}^2(r)} \right).
\end{equation}
with $v_j$ the  velocity relative to that of  the central galaxy along
axis  $j$,  and   $\sigma_{\rm  sat}(r)$  the  local,  one-dimensional
velocity dispersion obtained from solving the Jeans equation
\begin{equation}
\label{jeans}
\sigma^2_{\rm sat}(r) = {1 \over n_{\rm sat}(r)} 
\int_{r}^{\infty} n_{\rm sat}(r') {\partial \Psi \over \partial
  r}(r') {\rm d}r'
\end{equation}
with $\Psi(r)$ the gravitational  potential (Binney \& Tremaine 1987). 
Substituting   eq.~(\ref{nsatr})  for   the  spatial   number  density
distribution of satellites yields
\begin{eqnarray}
\label{siga1tr}
\sigma^2_{\rm sat}(r) &=& {c \, V^2_{\rm vir} \over 
{\cal R}^2 \mu_{1}(c)} \, 
\left({r \over {\cal R} r_s}\right)^{\alpha} \, 
\left(1 + {r \over {\cal R} r_s}\right)^{3-\alpha} \nonumber \\ 
 & & \int_{r/r_s}^{\infty} {\mu_1(x) {\rm d}x \over 
(x/{\cal R})^{\alpha+2} (1+x/{\cal R})^{3-\alpha}}
\end{eqnarray}
with 
\begin{equation}
\label{mudef}
\mu_{\alpha}(x) = \int_{0}^{x} y^{2-\alpha} \, (1+y)^{\alpha-3} \, {\rm d}y,
\end{equation}
For $\alpha  = {\cal R}  = 1$ the  satellite galaxies follow  the same
density  distribution as  the  dark matter  (i.e.,  no spatial  bias),
and~(\ref{siga1tr}) reduces to  the radial velocity dispersion profile
of a spherical NFW potential
\begin{equation}
\label{sig1dm}
\sigma^2_{\rm NFW}(r) =  {c \, V^2_{\rm vir} \over \mu_{1}(c)} \, 
\left({r \over r_s}\right) \, \left(1 + {r \over r_s}\right)^2 \, 
\int_{r/r_s}^{\infty} {\mu_1(x) {\rm d}x \over x^3 (1+x)^2}
\end{equation}
(cf.  Klypin \etal 1999).

\subsection{Creating Mock Surveys}
\label{sec:stack}

We  aim  to construct  MGRSs  with  the  same selection  criteria  and
observational biases  as in the 2dFGRS,  out to a  maximum redshift of
$z_{\rm max}=0.15$. We follow Yang \etal (2003b) and stack $4 \times 4
\times  4$ identical  $L_{300}$  boxes (which  have periodic  boundary
conditions),  and  place the  virtual  observer  in  the centre.   The
central $2  \times 2 \times  2$ boxes, are  replaced by a stack  of $6
\times 6 \times 6$ $L_{100}$ boxes  (see Fig.~11 in Yang \etal 2003b). 
This stacking geometry circumvents possible incompleteness problems in
the mock survey  due to insufficient mass resolution  of the $L_{300}$
simulations, and easily  allows us to reach $z_{\rm  max}=0.15$ in all
directions.   We   mimic  the  various   observational  selection  and
completeness effects in the 2dFGRS using the following steps:

\begin{enumerate}
  
\item We  define a $(\alpha,\delta)$-coordinate frame  with respect to
  the virtual observer at the centre of the stack of boxes, and remove
  all galaxies that are not located in the areas equivalent to the NGP
  and SGP regions of the 2dFGRS.
  
\item For each galaxy we compute the redshift as `seen' by the virtual
  observer.   We take  the observational  velocity  uncertainties into
  account  by   adding  a  random  velocity  drawn   from  a  Gaussian
  distribution  with  dispersion $85\kms$  (Colless  \etal 2001),  and
  remove those galaxies with $z > 0.15$.
 
\item For each  galaxy we compute the apparent  magnitude according to
  its luminosity and distance, to which we add a rms error of 0.15 mag
  (Colless  \etal  2001;  Norberg  \etal 2002b).   Galaxies  are  then
  selected  according  to   the  position-dependent  magnitude  limit,
  obtained using  the apparent magnitude  limit masks provided  by the
  2dFGRS team.
  
\item To take  account of the completeness level  of the 2dFGRS parent
  catalogue  (Norberg  \etal 2002b)  we  randomly  remove  9\% of  all
  galaxies.
  
\item  To  take  account  of  the  position-  and  magnitude-dependent
  completeness of the 2dFGRS, we randomly sample each galaxy using the
  completeness masks provided by the 2dFGRS team.
    
\end{enumerate}

Each MGRS  thus constructed  contains, on average,  $144000$ galaxies,
with a dispersion of $\sim 2600$ due to cosmic variance. As we show in
Section~\ref{sec:2dFres} this is in perfect agreement with the 2dFGRS.
In  addition, we  verified that  the MGRSs  also accurately  match the
clustering properties  (see Yang \etal 2003b),  the apparent magnitude
distribution  and  the redshift  distribution  of  the 2dFGRS.   Thus,
overall our MGRSs are fair representations of the 2dFGRS.
\begin{table}
\caption{Selection Criteria}
\begin{tabular}{lcccccc}
   \hline
SC & $R_h$ & $(\Delta V)_h$ & $f_h$ & $R_s$ & $(\Delta V)_s$ & $f_s$ \\
   & Mpc/h & km/s & & Mpc/h & km/s & \\
 (1) & (2) & (3) & (4) & (5) & (6) & (7) \\
\hline\hline
1 & $2.0$ & $1000$ & $2.0$ & $0.5$ & $1000$ & $4.0$ \\
2 & $2.0$ & $2000$ & $1.0$ & $0.5$ & $2000$ & $1.0$ \\
3 & $0.8 \sigma_{200}$ & $1000 \sigma_{200}$ & $1.0$ &  $0.15\sigma_{200}$ & $2000$ & $1.0$ \\
\hline
\end{tabular}
\medskip

\begin{minipage}{\hssize}
  Column~(1)  indicates  the  ID   of  the  selection  criterion,  the
  parameters of  which are listed in Columns~(2)  to~(7) and described
  in the  text.  $\sigma_{200}$ is defined as  the velocity dispersion
  of satellite galaxies around the host galaxy of interest in units of
  $200 \kms$.
\end{minipage}

\end{table}

\section{Methodology}
\label{sec:method}

\subsection{Selection Criteria}
\label{sec:select}

A galaxy is considered a potential host galaxy if it is at least $f_h$
times brighter than any other galaxy within a volume specified by $R_p
< R_h$ and  $\vert \Delta V \vert < (\Delta V)_h$.   Here $R_p$ is the
separation projected on the sky at the distance of the candidate host,
and $\Delta V$ is  the line-of-sight velocity difference.  Around each
potential  host  galaxy,  satellite  galaxies  are  defined  as  those
galaxies that  are at  least $f_s$ times  fainter than their  host and
located within a  volume with $R_p < R_s$ and $\vert  \Delta V \vert <
(\Delta V)_s$.  Host galaxies with zero satellite galaxies are removed
from the list of hosts.
\begin{figure*}
\centerline{\psfig{figure=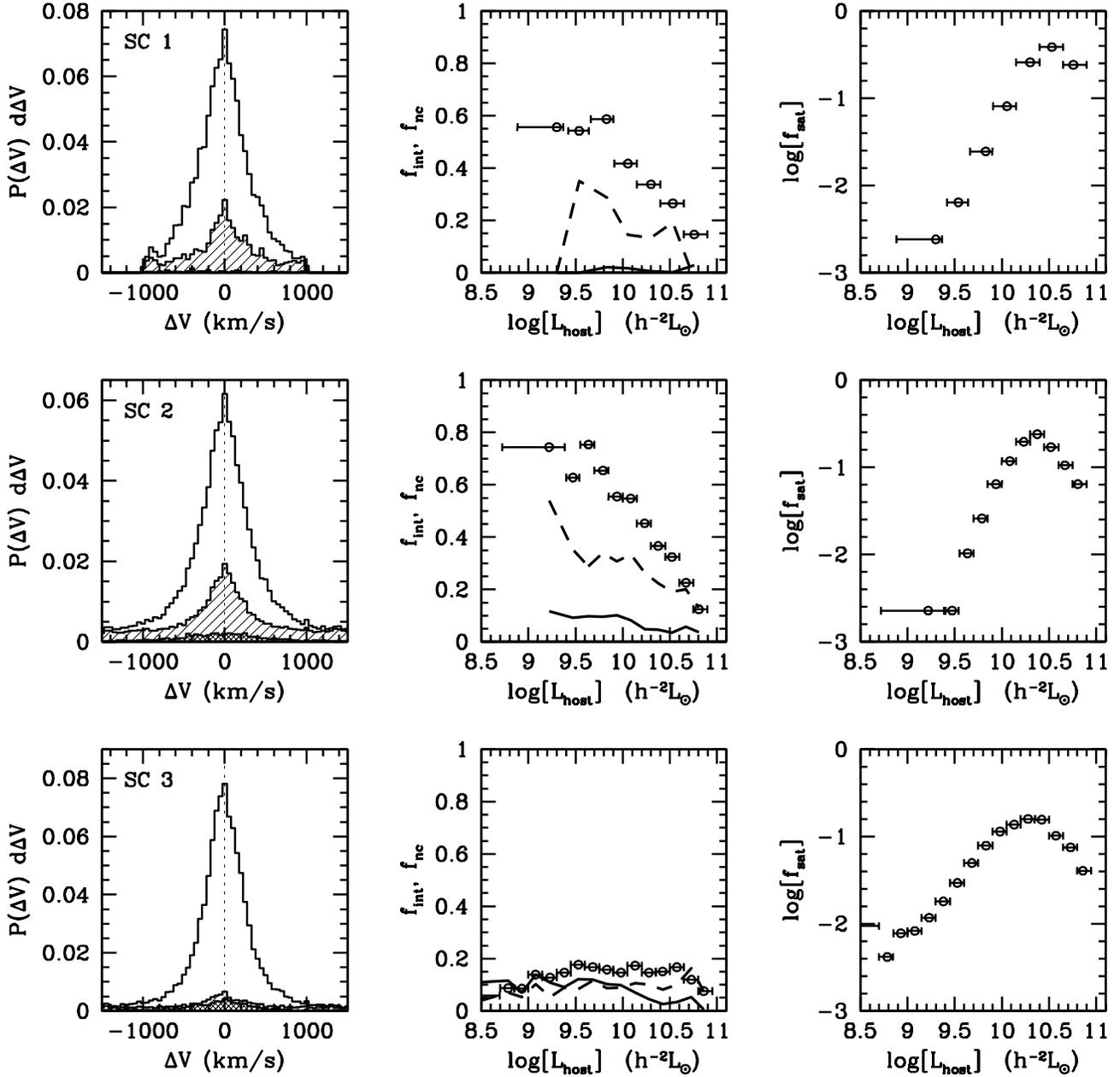,width=\hdsize}}
\caption{Various statistical properties of host/satellite pairs obtained
  from our MGRS using  three different selection criteria (SC). Panels
  on the  left plot  a histogram  of the distribution  of $\Delta  V =
  V_{\rm sat}  - V_{\rm host}$.  The contributions  of interlopers and
  satellites  around non-central  hosts are  indicated by  hatched and
  cross-hatched histograms, respectively.  Note that the $P(\Delta V)$
  of  interlopers  is not  uniform,  as  typically  assumed. The  open
  circles in the  panels in the middle column  indicate the interloper
  fraction  $f_{\rm int}$  for a  number of  bins in  $L_{\rm  host}$. 
  Horizontal  errorbars indicate  the range  of $L_{\rm  host}$  used. 
  Dashed lines  indicate the interloper fractions  obtained by fitting
  $P(\Delta   V)$    with   a   Gaussian   plus    a   constant   (see
  Section~\ref{sec:kinem}),  which  systematically  underpredicts  the
  true  interloper fraction.   Solid  lines indicate  the fraction  of
  satellites around  non-central host  galaxies.  Panels on  the right
  plot the fraction of satellites in each bin of $L_{\rm host}$.  Note
  how  the  number of  satellites  decreases  rapidly with  decreasing
  $L_{\rm host}$,  and that this  decrease is less pronounced  with SC
  3.}
\label{fig:sc}
\end{figure*}

In total,  the selection of hosts  and satellites thus  depends on six
free  parameters:  $R_h$, $(\Delta  V)_h$  and  $f_h$  to specify  the
population of  host galaxies, and  $R_s$, $(\Delta V)_s$ and  $f_s$ to
specify the  satellite galaxies.  These parameters also  determine the
number of  interlopers (defined as a galaxy  not physically associated
with the halo of the host  galaxy) and non-central hosts (defined as a
host  galaxy that  is not  the brightest,  central galaxy  in  its own
halo).   Minimizing the  number of  interlopers  requires sufficiently
small $R_s$  and $(\Delta V)_s$. Minimizing the  number of non-central
hosts  requires  one  to   choose  $R_h$,  $(\Delta  V)_h$  and  $f_h$
sufficiently large.  Of course, each of these restrictions dramatically
reduces  the   number  of  both  hosts  and   satellites,  making  the
statistical estimates more and more noisy.

We  compare three  different selection  criteria (hereafter  SC), that
only differ  in their  values for the  six parameters  described above
(see Table~1), and use our MGRS to investigate the resulting fractions
of interlopers and non-central  hosts.  Most results are summarized in
Table~2  and Fig.~\ref{fig:sc}.   SC 1  is identical  to that  used by
McKay \etal  (2002) and Brainerd  \& Specian (2003),  who investigated
the  kinematics  of  satellite   galaxies  in  the  SDSS  and  2dFGRS,
respectively.  The  same SC  was also used  by Prada \etal  (2003) for
their sample 3.  It uses  a fairly restrictive set of parameters; host
galaxies must be  at least two times more  luminous ($f_h=2$) than any
other galaxy  within a volume specified  by $R_h = 2  h^{-1} \Mpc$ and
$(\Delta V)_h = 1000 \kms$.   Satellite galaxies are selected as those
galaxies within $R_s = 0.5 h^{-1} \Mpc$ and $(\Delta V)_s = 1000 \kms$
around each  host that are at  least four times fainter  ($f_s = 4.0$)
than the  host.  Applying these  SC to our  MGRS (which consists  of a
total of 143727 galaxies), yields 1851 hosts and 3876 satellites.  The
fraction of interlopers  is 27 percent, while only  one percent of the
hosts is non-central (see Table~2).

The open circles in the  upper, middle panel of Fig.~\ref{fig:sc} show
that  the  interloper  fraction,   $f_{\rm  int}$,  is  larger  around
lower-luminosity hosts,  reaching as high as $\sim  60$ percent around
the  faintest hosts in  the sample.   Clearly, accurate  estimates for
halo masses based  on the kinematics of satellite  galaxies requires a
proper correction for these interlopers, and thus a detailed knowledge
of their velocity distribution, $P_{\rm int}(\Delta V)$. Thus far, the
standard approach has  been to assume that $P_{\rm  int}(\Delta V)$ is
uniform (i.e., McKay \etal 2002; Brainerd \& Specian 2003; Prada \etal
2003).  The upper left-hand panel of Fig.~\ref{fig:sc} plots $P(\Delta
V)$ for host-satellite  pairs selected from our MGRS  using SC 1.  The
hatched  histogram shows  the contribution  of  interlopers.  Clearly,
$P_{\rm  int}(\Delta  V)$  is  not  uniform,  but  instead  reveals  a
pronounced peak  around $\Delta V = 0$.   This is due to  (i) the fact
that  galaxies, including  interlopers,  are clustered,  and (ii)  the
infall  of  galaxies  around  overdense  regions. As  we  show  below,
assuming  a  uniform  $P_{\rm  int}(\Delta  V)$  in  the  analysis  of
satellite kinematics  (as is generally  done) results in  a systematic
underestimate  of $f_{\rm  int}$ but,  fortunately, does  not  lead to
significant errors in the kinematics.
  
With SC  2 our main objective is  to increase the number  of hosts and
satellites, and  to include  galaxy groups and  clusters on  a similar
footing as `isolated'  galaxies.  In SC 2 we  therefore set both $f_h$
and $f_s$ to  unity.  This greatly increases the  number of both hosts
and satellites,  and allows brightest cluster galaxies  to be included
as host  galaxies.  In order to  cover a sufficiently  large volume in
velocity  space  to properly  sample  rich  clusters  we enlarge  both
$(\Delta V)_h$ and  $(\Delta V)_s$ to $2000 \kms$.   As we show below,
this   has  the  additional   advantage  that   it  allows   a  better
determination of the contribution of interlopers, and therefore a more
accurate, statistical correction. SC 2 results in 4 to 5 times as many
hosts  ($N_{\rm host}=7863$) and  satellites ($N_{\rm  sat}=19099$) as
with SC 1.   However, the fraction of interlopers  has also increased,
from  27  to  39  percent.    Fortunately,  as  is  evident  from  the
middle-left   panel  of  Fig.~\ref{fig:sc},   most  of   these  excess
interlopers have $1000 \kms \leq  \vert \Delta V \vert \leq 2000 \kms$
and are easily corrected  for (see Section~\ref{sec:kinem}). As for SC
1,  the fraction  of  interlopers increases  strongly with  decreasing
$L_{\rm  host}$, with  more  than 70  percent  interlopers around  the
faintest  hosts in  the sample.   The fraction  of  non-central hosts,
$f_{\rm nc}$, has increased from one to five percent compared to SC 1.
This is mainly  a consequence of setting $f_h=1$,  which allows bright
galaxies in the outskirts of  clusters to be (erroneously) selected as
hosts.  As evident from the cross-hatched histogram in the middle-left
panel of  Fig.~\ref{fig:sc}, the  $P(\Delta V)$ of  satellite galaxies
around these non-central hosts is very broad.  As for the interlopers,
accurate  estimates  of  halo  masses based  on  satellite  kinematics
require keeping the number of non-central hosts as small as possible.
\begin{table*}
\caption{Satellite Kinematics.}
\begin{tabular}{lcccccccccc}
   \hline
Survey & SC & $N_{\rm total}$ &
$N_{\rm host}$ & $N_{\rm sat}$ & $f_{\rm nc}$ & $f_{\rm int}$ & 
$f_{\rm int}^{\rm fit}$ & $\sigma_{10}$ & $a_1$ & $a_2$ \\
 (1) & (2) & (3) & (4) & (5) & (6) & (7) & 
(8) & (9) & (10) & (11) \\
\hline\hline
MGRS   & 1 & $143727$ &  $1851$ &  $3876$ & $0.01$ & $0.27$ & $0.20$ & $130$ & $0.55$ &  $--$ \\
MGRS   & 2 & $143727$ &  $7863$ & $19099$ & $0.05$ & $0.39$ & $0.28$ & $185$ & $0.37$ & $0.17$\\
MGRS   & 3 & $143727$ & $10483$ & $16750$ & $0.07$ & $0.15$ & $0.12$ & $176$ & $0.52$ & $0.11$\\
2dFGRS & 3 & $146735$ &  $8132$ & $12569$ &  $--$  &  $--$  & $0.19$ & $193$ & $0.48$ & $0.13$\\
\hline
\end{tabular}
\medskip

\begin{minipage}{\hdsize}
  Column~(1)  indicates  whether  the  sample of  host  and  satellite
  galaxies  has been  extracted  from our  MGRS  or from  the 2dFGRS.  
  Column~(2) indicates the selection criterion used, the parameters of
  which are listed in  Table~1. Columns~(3) indicates the total number
  of galaxies  in the survey,  while columns~(4) and~(5)  indicate the
  numbers of  host and satellite  galaxies, respectively.  Columns~(6)
  and~(7) indicate the fraction  $f_{\rm nc}$ of non-central hosts and
  the (true) fraction of interlopers  $f_{\rm int}$, both of which are
  only  known for  the  MGRS. Finally,  columns~(8)  to~(11) list  the
  best-fit  parameters  obtained from  the  maximum likelihood  method
  described in  Section~\ref{sec:kinem}.
\end{minipage}

\end{table*}

Both selection  criteria discussed so far yield  fairly high fractions
of  interlopers, especially around  faint hosts.   In addition,  as is
evident from  the panels on the right-hand  side of Fig.~\ref{fig:sc},
the number of host-satellite  pairs decreases very rapidly for $L_{\rm
  host} \lta 3 \times 10^{10} h^{-2} \Lsun$. This is mainly due to the
limited  number of  faint  hosts  that make  the  selection criteria.  
Ideally, one would use adaptive selection criteria, adjusting $(\Delta
V)_h$, $R_h$ and $R_s$ to the virial radius and virial velocity of the
halo of  the host galaxy  in question.  This, however,  requires prior
knowledge of the halo masses as a function of $L_{\rm host}$, which is
exactly what we  are trying to recover from  the satellite kinematics. 
We therefore use  an iterative procedure: start from  an initial guess
for  $\sigma_{\rm sat}(L_{\rm host})$  and estimate  the corresponding
virial  radius and velocity  around each  individual host  galaxy. Use
these to adapt $(\Delta V)_h$, $R_h$,  and $R_s$ to the host galaxy in
question, and select a new sample of host-satellite pairs. Use the new
sample  to obtain  an  improved estimate  of $\sigma_{\rm  sat}(L_{\rm
  host})$,  and start  the next  iteration.  In  detail we  proceed as
follows:
\begin{enumerate}
  
\item  Use SC 2 to select hosts and satellites
  
\item  Fit the  satellite kinematics  of the  resulting sample  with a
  simple functional form (see Section~\ref{sec:kinem}).
  
\item  Select new  hosts and  satellites  using $(\Delta  V)_h =  1000
  \sigma_{200}  \kms$,  $(\Delta  V)_s   =  2000  \kms$,  $R_h  =  0.8
  \sigma_{200}  h^{-1}  \Mpc$, and  $R_s  =  0.15 \sigma_{200}  h^{-1}
  \Mpc$.  Here  $\sigma_{200}$ is $\sigma_{\rm  sat}(L_{\rm host})$ in
  units of $200 \kms$.
  
\item Repeat (ii) and (iii) until $\sigma_{\rm sat}(L_{\rm host})$ has
  converged to the required precision.  Typically this requires 3 to 4
  iterations.

\end{enumerate}
The numerical values  in step~(iii) are based on  extensive tests with
our MGRSs, optimizing the  results using the following criteria: large
$N_{\rm  sat}$ and  $N_{\rm host}$,  small $f_{\rm  int}$  and $f_{\rm
  nc}$,  and good  sampling of  $L_{\rm host}$.   The $R_h$  and $R_s$
correspond  roughly to  $2.0$  and $0.375$  times  the virial  radius,
respectively.   Applying this  adaptive SC  to our  MGRS  yields 10483
hosts and 16750  satellites (after 4 iterations).  The  number of host
galaxies has drastically  increased with respect to SC  2.  As we show
below, this allows  us to probe the satellite  kinematics down to host
galaxies  with much  fainter  luminosities.  The  number of  satellite
galaxies, on the other hand, has decreased with respect to SC 2.  This
mainly reflects a drastic decrease  in the number of interlopers, from
39 to 15 percent.  More importantly, the interloper fraction no longer
strongly depends on $L_{\rm host}$.

\subsection{Analytical Estimates}
\label{sec:analest}

When investigating the impact  of interlopers and non-central hosts on
the  kinematics   of  satellite  galaxies,   and  comparing  different
selection criteria, it is useful to have an analytical estimate of the
expected  $\sigma_{\rm  sat}(L)$.   This  section  describes  how  the
conditional luminosity function (CLF)  may be used to compute $\langle
\sigma_{\rm  sat}(L) \rangle$ for  a flux-limited  survey such  as the
2dFGRS.

For  a halo  of  mass $M$,  the  expectation value  for the  projected
velocity dispersion of satellite galaxies is given by
\begin{equation}
\label{sigexp}
\langle \sigma_{\rm sat} \rangle_M = 
{4 \pi \over \langle N_{\rm sat} \rangle_M}  \int_{0}^{r_{\rm vir}}
n_{\rm sat}(r) \, \sigma_{\rm sat}(r) \, r^2 \, {\rm d}r
\end{equation}
with  $\langle N_{\rm sat}  \rangle_M$ the  mean number  of satellites
with $L \geq L_{\rm min}$ in a halo of mass $M$ which is given by
\begin{eqnarray}
\label{nsat}
\langle N_{\rm sat} \rangle_M & = &
\int_{L_{\rm min}}^{\infty} \Phi(L \vert M) \, {\rm d}L - 1 \nonumber \\
& \equiv & 4 \pi \int_{0}^{r_{\rm vir}} n_{\rm sat}(r) \, r^2 \, {\rm d}r
\end{eqnarray}
Substituting~(\ref{nsatr}) and~(\ref{siga1tr}) yields
\begin{equation}
\label{avsigtr}
\langle \sigma_{\rm sat} \rangle_M = {V_{\rm vir}
\over \mu_{\alpha}(c/{\cal R})} \, \sqrt{c \over {\cal R} \, \mu_1(c)} \,
\int_{0}^{c/{\cal R}} {y^{2-\alpha/2} \, {\cal I}^{1/2}(y) \over 
(1+y)^{(3-\alpha)/2}} \, {\rm d}y
\end{equation}
with
\begin{equation}
\label{cali}
{\cal I}(y) = \int_{y}^{\infty} 
{\mu_1({\cal R} \tau) \,{\rm d}\tau \over \tau^{\alpha+2} (1+\tau)^{3-\alpha}}
\end{equation}

In  general, there  will not  be a  one-to-one,  purely deterministic,
relation  between  halo mass  and  host  luminosity.  Therefore,  when
averaging over all host  galaxies of given luminosity, the expectation
value for the velocity dispersion of their satellite galaxies is given
by
\begin{equation}
\label{expsighost}
\langle \sigma_{\rm sat}(L_c) \rangle = \int_0^{\infty}  
P(M \vert L_c) \, \langle \sigma_{\rm sat} \rangle_M \, {\rm d}M 
\end{equation}
with $P(M \vert L_c) {\rm d}M$ the conditional probability that a {\it
  central} galaxy with  luminosity $L_c$ resides in a  halo of mass $M
\pm  {\rm d}M/2$,  and  $\langle \sigma_{\rm  sat}  \rangle_M$ is  the
expectation value for the  projected velocity dispersion of satellites
in  a halo  of  mass  $M$ given  by  eq.~(\ref{avsigtr}). Using  Bayes
theorem, we rewrite~(\ref{expsighost}) as
\begin{equation}
\label{expsigbayes}
\langle \sigma_{\rm sat}(L_c) \rangle = {\int_0^{\infty}  
P(L_c \vert M) \, n(M) \, \langle \sigma_{\rm sat} \rangle_M \, {\rm d}M 
\over \int_0^{\infty} P(L_c \vert M) \, n(M) \, {\rm d}M}
\end{equation}
with  $n(M)$  the  halo  mass   function  and  $P(L_c  \vert  M)$  the
conditional probability that a halo of mass $M$ hosts a central galaxy
with luminosity $L_c$.  In our  MGRS, $L_c$ is drawn `constrained' for
which
\begin{equation}
\label{casedet}
P(L_c \vert M) = \left\{ \begin{array}{ll}
\Phi(L_c \vert M) & \mbox{if $L_c \geq L_1(M)$} \\
0                 & \mbox{if $L_c < L_1(M)$}
\end{array} \right.
\end{equation}
with  $L_1(M)$ defined by  eq.~(\ref{Lj}). In  Appendix~B we  show how
$P(L_c \vert  M)$ can  be obtained  in the case  where $L_c$  is drawn
`randomly'  (as opposed to  `constrained') from  the CLF.   The dashed
curve  in  Fig~\ref{fig:analest}   plots  the  host-averaged  $\langle
\sigma_{\rm  sat}(L_c)  \rangle$  thus  obtained (see  Appendix~A  for
details regarding the CLF  used).  This $\langle \sigma_{\rm sat}(L_c)
\rangle$ is  the true mean  velocity dispersion of  satellite galaxies
around hosts with  luminosity $L_c$, where the mean  is taken over the
number of host galaxies.

Unfortunately,  this   is  not  what  an  observer   who  stacks  many
host-satellite  pairs  together  obtains,  which  instead  is  a  {\it
  satellite-weighted}  mean.   Since  more  massive  haloes  typically
contain  more  satellites,  satellite  weighting  will  bias  $\langle
\sigma_{\rm sat}(L_c) \rangle$ high  with respect to the host-weighted
mean. We  can use the  CLF to estimate  the magnitude of this  `bias'. 
The satellite-weighted expectation  value for $\sigma_{\rm sat}(L_c)$,
in  a  {\it  volume-limited}   survey  complete  down  to  a  limiting
luminosity of $L_{\rm min}$ is
\begin{equation}
\label{expsig}
\langle \sigma_{\rm sat}(L_c) \rangle =
{\int_0^{\infty}  P(M \vert L_c) \, 
\langle N_{\rm sat} \rangle_M \, \langle \sigma_{\rm sat} \rangle_M
\, {\rm d}M \over 
\int_0^{\infty}  P(M \vert L_c) \, \langle N_{\rm sat} 
\rangle_M \, {\rm d}M}
\end{equation}
The   dotted  line  in   Fig.~\ref{fig:analest}  shows   the  $\langle
\sigma_{\rm sat}(L_c) \rangle$ thus  obtained.  For $L_{\rm host} \gta
3  \times  10^9  h^{-2}  \Lsun$  the satellite-weighted  mean  is,  as
expected, larger than the host-weighted  mean (by as much as $\sim 40$
percent).  For  less luminous hosts, however,  satellite averaging has
no significant effect.  This is due to the detailed functional form of
$\langle N_{\rm  sat} \rangle_M$, which  is much shallower at  low $M$
than at high $M$ (e.g., van den Bosch \etal 2003a).

The above  estimate is based on  a complete sampling  of the satellite
population of each  halo.  In reality, however, there  are two effects
that result in  a reduced completeness.  First of  all, satellites are
only  selected within  a certain  projected  radius around  the host.  
Whenever  that  radius is  smaller  than  the  virial radius,  only  a
fraction of all satellites enter  the sample, and with a mean velocity
dispersion   that  differs   from~(\ref{avsigtr}).   Secondly,   in  a
flux-limited survey  the number of  satellites around a host  of given
luminosity depends on  redshift.  Again, we can account  for these two
effects using simple algebra.

Suppose  we observe a  host-satellite system  in projection  through a
circular  aperture with radius  $R_p$. The  expectation value  for the
observed velocity dispersion  of the satellites in a  halo of mass $M$
is given by
\begin{equation}
\label{sigproj}
\langle \sigma_{\rm sat} \rangle_M = {
\int_0^{R_p} {\rm d}R \, R \int_{R}^{r_{\rm vir}} 
{r {\rm d}r \over \sqrt{r^2 - R^2}} \, n_{\rm sat}(r) \, \sigma_{\rm sat}(r)
\over
\int_0^{R_p} {\rm d}R \, R \int_{R}^{r_{\rm vir}} 
{r {\rm d}r \over \sqrt{r^2 - R^2}} \, n_{\rm sat}(r)}
\end{equation}
which  is straightforward  to compute  upon substituting~(\ref{nsatr})
and~(\ref{siga1tr}).  For  $R_p = r_{\rm  vir}$ one obtains  the total
projected velocity given by~(\ref{avsigtr}).

The  expectation value for  a {\it  flux-limited} survey  follows from
integrating~(\ref{expsigbayes}) over redshift:
\begin{equation}
\label{expsigflux}
\langle \sigma_{\rm sat}(L_c) \rangle = {1 \over V}
\int_0^{\Omega} {\rm d}\Omega \, \int_0^{z_{\rm max}} {\rm d}z 
{{\rm d}V \over {\rm d}\Omega \, {\rm d}z} \, 
\langle \sigma_{\rm sat}(L_c,z) \rangle
\end{equation}
Here $\Omega$ is  the solid angle of sky of the  survey, ${\rm d}V$ is
the differential volume  element, and $z_{\rm max}$ is  the minimum of
the survey redshift limit (0.15 in our MGRSs) and the maximum redshift
out to which a galaxy with  luminosity $L_c$ can be detected given the
apparent  magnitude  limit  of  the  survey.   The  expectation  value
$\langle      \sigma_{\rm       sat}(L_c,z)      \rangle$      follows
from~(\ref{expsigbayes})  upon replacing  $n(M)$ with  $n(M,z)$ (i.e.,
accounting for the evolution  in the halo mass function)\footnote{Note
  that we assume  here that the CLF does not  evolve with redshift, at
  least  not  over  the  small  range  of  redshift  ($z  \leq  0.15$)
  considered here.}.   In addition, since the minimum  luminosity of a
galaxy in a flux-limited survey  depends on redshift the $L_{\rm min}$
in~(\ref{nsat}) needs to be  replaced with $L_{\rm min}(z)$.  Finally,
in case  $R_p < r_{\rm vir}$  one needs to  use $\langle \sigma^2_{\rm
  sat}      \rangle_M$       given      by~(\ref{sigproj})      rather
than~(\ref{avsigtr}).
\begin{figure}
\centerline{\psfig{figure=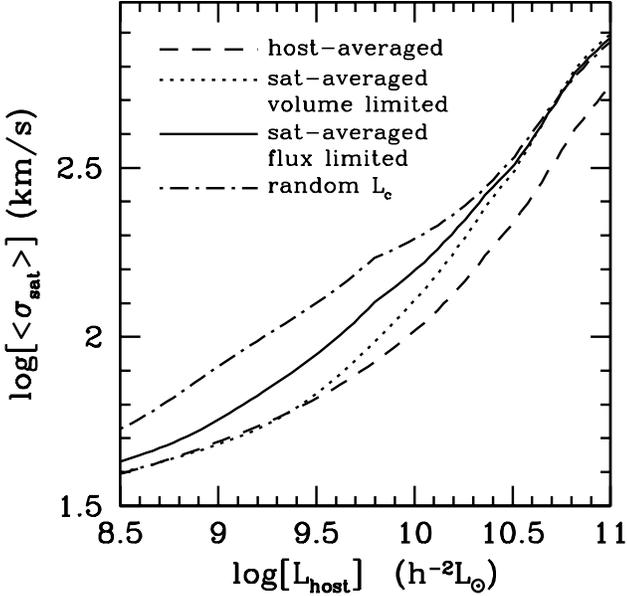,width=\hssize}}
\caption{Expectation values for $\sigma_{\rm sat}$ as a function of
  $L_{\rm host}$ computed using the  CLF as explained in the text. The
  dashed line corresponds to  the true, host-averaged mean. The dotted
  line shows  the satellite-weighted mean,  which is what  an observer
  obtains  from stacking  many  host-satellite pairs  (using a  volume
  limited  survey).    The  solid   and  dot-dashed  lines   show  the
  expectation  values  for   a  flux-limited  survey,  computed  using
  `constrained' and `random' host luminosities, respectively. See text
  for details.}
\label{fig:analest}
\end{figure}

The   solid  curve  in   Fig.~\ref{fig:analest}  shows   the  $\langle
\sigma_{\rm sat}(L_c) \rangle$ thus obtained with $R_p = r_{\rm vir}$.
Overall,  the  expectation  value  for $\sigma_{\rm  sat}(L_c)$  of  a
flux-limited survey  is larger than for a  volume-limited survey. This
owes to the fact that, because  of the flux limit, smaller mass haloes
loose  a  relatively  larger  fraction of  satellites.   Finally,  the
dot-dashed line  shows the same  expectation value, satellite-averaged
over  a flux-limited  survey, but  computed assuming  `random', rather
than  `constrained',  luminosities  for  the central  galaxies.   This
increases the  width of the conditional  probability distribution $P(M
\vert  L_c)$  (see  Appendix~B),  which  in  turn  strongly  increases
$\langle \sigma_{\rm sat}(L_c) \rangle$.

Clearly,  satellite  weighting  from flux-limited  surveys  introduces
large systematic  biases in the kinematics of  satellite galaxies. The
magnitude of this  bias depends on, among others,  host luminosity and
the second moment of $P(M \vert L_c)$, and can easily be as large as a
factor  two. Since, to  first order,  $M \propto  \sigma_{\rm sat}^3$,
this implies  a systematic  overestimate of halo  masses of  almost an
order  of magnitude. Clearly,  if one  were not  to correct  for these
systematic  biases, the mass-to-light  ratios inferred  from satellite
kinematics  are   systematically  too  high   by  the  same   amount.  
Unfortunately, such  bias correction is  model-dependent.  Although it
is  straightforward to  correct  for  the bias  due  to the  satellite
averaging in a  {\it volume limited survey} (by  simply weighting each
satellite  by the  inverse  of  the number  of  satellites around  the
corresponding  host  galaxy), in  a  flux-limited  survey  one has  to
correct  for {\it  missing} satellites  (those that  did not  make the
flux-limit).   This requires  prior  knowledge of  the abundances  and
luminosities of  satellite galaxies, and  is thus model  dependent. As
shown here,  the CLF formalism is one  such model that can  be used to
model these biases in a straightforward way.
\begin{figure*}
\centerline{\psfig{figure=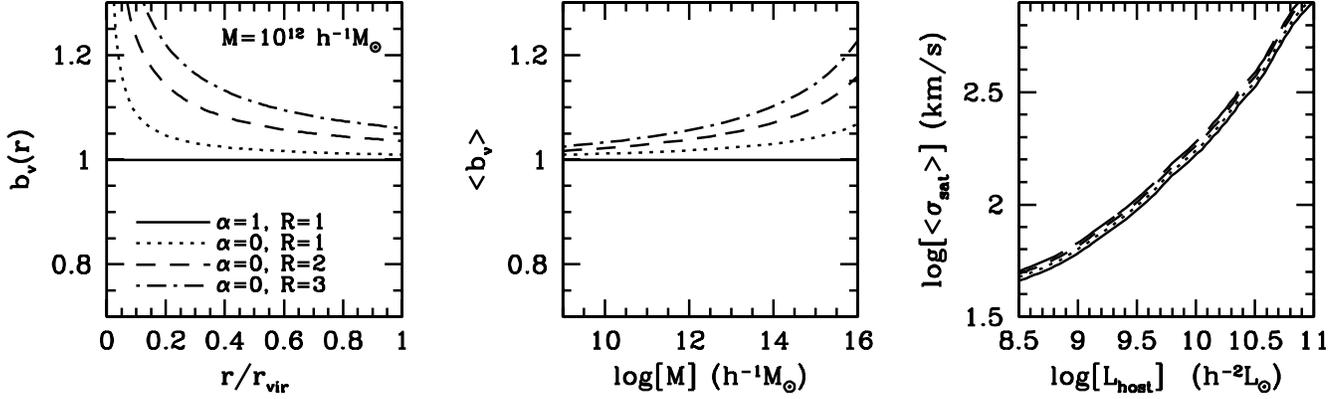,width=\hdsize}}
\caption{The left-hand panel plots the {\it local} velocity bias $b_v(r)$  
  of  isotropic, steady-state populations  of satellite  galaxies that
  are  spatially anti-biased with  respect to  the mass  distribution. 
  Results are  shown as a function  of radius (normalized  to the halo
  virial  radius)  for   three  different  spatial  distributions,  as
  indicated. For  comparison, the solid  line corresponds to  the case
  without spatial bias.  Note how spatial anti-bias induces a positive
  velocity bias.  The panel in the middle shows the corresponding {\it
    global}  velocity bias, $\langle  b_v \rangle$,  as a  function of
  halo mass.   More massive haloes reveal a  larger, positive velocity
  bias.  Finally, the right-hand panel plots corresponding expectation
  values for $\sigma_{\rm sat}(L_{\rm host}$).  The larger the spatial
  anti-bias, the higher the expectation values, although the effect is
  marginal.}
\label{fig:vbias}
\end{figure*}

\subsubsection{The connection between spatial bias and velocity bias}
\label{sec:bias}

The expectation  values discussed so  far are based on  the assumption
that  satellite galaxies  follow  a NFW  number density  distribution,
i.e., we  assumed a  constant mass-to-number density  ratio.  Numerous
studies have  shown this  to be a  good approximation for  clusters of
galaxies (e.g., Carlberg  \etal 1997a; van der Marel  \etal 2000; Lin,
Mohr  \&  Stanford 2004;  Rhines  \etal  2004;  Diemand \etal  2004).  
However, numerical  simulation have  shown that dark  matter subhaloes
typically  are   spatially  anti-biased  with  respect   to  the  mass
distribution (i.e.,  Moore \etal 1998, Ghigna \etal  1999; Colin \etal
1999; Klypin \etal  1999; Okamoto \& Habbe 1999;  Springel \etal 2001;
De  Lucia \etal  2004;  Diemand \etal  2004).   If satellite  galaxies
follow a similar  anti-bias this will reflect itself  on the dynamics. 
We  can  model spatial  anti-bias  by  adjusting  the free  parameters
$\alpha$ and ${\cal R}$.   Setting $\alpha=0$, for example, introduces
a constant  number density core,  and thus spatial anti-bias  at small
radii. The parameter ${\cal R}$ gives  the ratio of the scale radii of
satellite  galaxies and dark  matter, and  can be  used to  adjust the
radius of the core region.

Fig.~\ref{fig:vbias} illustrates  the effect that spatial  bias has on
the dynamics of satellite galaxies. We define the {\it local} velocity
bias as
\begin{equation}
\label{vbiastrue}
b_v(r) \equiv \left[ {\sigma^2_{\rm sat}(r) \over \sigma^2_{\rm dm}(r)} 
\right]^{1/2}
\end{equation}
with   $\sigma^2_{\rm   sat}(r)$   and  $\sigma^2_{\rm   dm}(r)$   the
one-dimensional, isotropic velocity dispersions of satellites and dark
matter  particles,  given  by eq.~(\ref{siga1tr})  and~(\ref{sig1dm}),
respectively.   If $b_v >  1$ then  satellite galaxies  typically move
faster than  dark matter particles in  the same halo, and  we speak of
positive velocity bias.   If $b_v < 1$ the  satellites are dynamically
colder than the  dark matter particles, and the  velocity bias is said
to  be  negative.   In  addition  to the  {\it  local}  velocity  bias
$b_v(r)$, we also define the {\it global} velocity bias
\begin{equation}
\label{vbias}
\langle b_v \rangle \equiv \left[ {\langle \sigma^2 \rangle_{\rm sat} 
\over \langle \sigma^2 \rangle_{\rm dm}} \right]^{1/2}
\end{equation}
where  $\langle  \cdot  \rangle$  indicates  mass  or  number-averaged
quantities   (cf.   eq.~[\ref{sigexp}]).    The  left-hand   panel  of
Fig.~\ref{fig:vbias} plots, for a  halo with $M=10^{12} h^{-1} \Msun$,
the local  velocity bias $b_v(r)$  for three different  populations of
satellite galaxies with $\alpha=0$ and ${\cal R} = 1$, $2$, and $3$ as
indicated.   Note the  overall positive  velocity bias,  which reaches
very  high  values at  small  radii:  a  (dynamically relaxed)  tracer
population  that   is  less  centrally  concentrated   than  the  mass
distribution  has a  positive velocity  bias (see  also  Diemand \etal
2004).  The middle panel of Fig.~\ref{fig:vbias} shows that the global
velocity bias $\langle b_v \rangle$  is larger for more massive haloes
(since these  have smaller  halo concentrations $c$).   The right-hand
panel, finally, plots the expectation values $\langle \sigma_{\rm sat}
\rangle$ as a function of $L_{\rm host}$.  In the case of $\alpha = 0$
and ${\cal R} = 3$ the  expectation values are a factor $1.1$ to $1.2$
larger than for  the case without spatial bias  ($\alpha={\cal R}=1$). 
Thus, the  effect of spatial (anti)-bias  is fairly small;  it is much
more  important   to  have  accurate  knowledge   of  the  conditional
probability function $P(M \vert L_c)$  than of $n_{\rm sat}(r)$ if one
is to obtain accurate, unbiased estimates of $\langle \sigma_{\rm sat}
\rangle$ as a function of $L_{\rm host}$.

\subsubsection{The impact of orbital anisotropies}
\label{sec:anis}

So  far we  have assumed  that the  orbits of  satellite  galaxies are
isotropic. However, numerical simulations  indicate that the orbits of
dark matter  subhaloes, although close  to isotropic near  the centre,
become  slightly  radially anisotropic  at  larger halo-centric  radii
(e.g.   Diemand  \etal  2004).    In  this  section  we  estimate  how
anisotropy impacts  on the projected velocity  dispersion of satellite
galaxies.

The  expectation  value  for  the  projected  velocity  dispersion  of
satellite galaxies  is given by  eq.~(\ref{sigexp}).  As long  as this
expectation  value  integrates  over  all  satellites,  it  is  almost
independent of the anisotropy of  the orbits. This is most easily seen
by considering the virial theorem,  which states that for a virialized
system $\langle \sigma^2 \rangle_M = \vert  W \vert / M$, with $W$ the
system's  total potential  energy. Thus,  as long  as  $\langle \sigma
\rangle  \sim  \langle  \sigma^2   \rangle^{1/2}$,  which  is  a  good
approximation  for  most  realistic  systems, the  projected  velocity
dispersion of  satellite galaxies  should be virtually  independent of
the anisotropy of their orbits.

Contrary to the {\it global}  value of $\langle \sigma^2 \rangle$, the
{\it local} velocity dispersion  depends quite strongly on anisotropy. 
Therefore, as soon as one only considers a radially dependent fraction
of the satellites, as is the case with our selection criterion 3 where
we integrate over  a circular aperture with radius  $R_{\rm ap} \simeq
0.375 R_{\rm vir}$, the impact  of anisotropy is no longer necessarily
negligible. In order to estimate the amplitude of this effect we first
solve the Jeans equation in spherical symmetry
\begin{equation}
\label{jeanseq}
{{\rm d} \over {\rm d}r} \rho \sigma_r^2 + {2 \beta \over r} \rho
\sigma_r^2 + \rho {{\rm d}\Psi \over {\rm d}r} = 0
\end{equation}
(Binney \& Tremaine 1987).  Here  $\beta(r) = 1 - {\sigma_t^2(r) \over
  2  \sigma_r^2(r)}$  is a  measure  of  the  orbital anisotropy,  and
$\sigma_r(r)$ and  $\sigma_t(r)$ are  the velocity dispersions  in the
radial  and tangential directions,  respectively. Assuming  a constant
value  for   $\beta$,  and  using  the   boundary  condition  $\rho(r)
\sigma^2(r) \rightarrow 0$ for $r \rightarrow \infty$, the solution to
this linear differential equation of first order is
\begin{equation}
\label{jeanseqsol}
\rho(r) \sigma_r^2(r) = {G \over r^{2 \beta}} 
\int_{r}^{\infty} r^{2 \beta-2} \, \rho(r) \, M(r) \, {\rm d}r
\end{equation}
with  $G$ the  gravitational  constant and  $M(r)$  the mass  enclosed
within radius  $r$. The expectation  value for the  projected velocity
dispersion  integrated over  a circular  aperture with  radius $R_{\rm
  ap}$ is given by
\begin{equation}
\label{sigprojap}
\langle \sigma^2 \rangle(\beta) = {\int_{0}^{R_{\rm ap}} {\rm d}R \, R \, 
\int_{R}^{R_{\rm vir}} \rho(r) \sigma_r^2(r) 
{1 - \beta (R/r)^2 \over \sqrt{1 - (R/r)^2}} {\rm d}r \over
\int_{0}^{R_{\rm ap}} {\rm d}R \, R \, 
\int_{R}^{R_{\rm vir}} {\rho(r) \over \sqrt{1 - (R/r)^2}} {\rm d}r}
\end{equation}
where we have  made use of eq~[4-60] in Binney  \& Tremaine (1987). To
compute   the    impact   of   orbital    anisotropy   we   substitute
eq.~(\ref{jeanseqsol}) in the above  expression and use an NFW density
distribution to compute the ratio
\begin{equation}
\label{sigrat}
\Sigma(\beta) = \left[{\langle \sigma^2(\beta)\rangle \over
\langle \sigma^2(0)\rangle}\right]^{1/2}
\end{equation}
where we set  $R_{\rm ap} = 0.375 R_{\rm vir}$  as appropriate for our
selection criterion 3. We find  that $\Sigma$ increases from $0.97$ to
$1.05$ when $\beta$ increases from $-0.5$ to $0.5$. Clearly, even when
using an aperture  that is only about one-third  of the virial radius,
orbital  anisotropy  effects  the  projected  velocity  dispersion  of
satellite galaxies only at the level of a few percent.

\subsection{Kinematics}
\label{sec:kinem}

The expectation values  discussed above are all based  on an idealized
situation without  interlopers and non-central hosts.  We  now turn to
our  MGRSs  from  which  we  select hosts  and  satellites  using  the
selection criteria discussed  in Section~\ref{sec:select}.  We analyze
the kinematics  of the satellite  galaxies and compare the  results to
the  analytical  predictions  presented  above.   This  allows  us  to
investigate the  impact of interlopers  and non-central hosts,  and to
properly compare the different selection criteria.

\begin{figure*}
\centerline{\psfig{figure=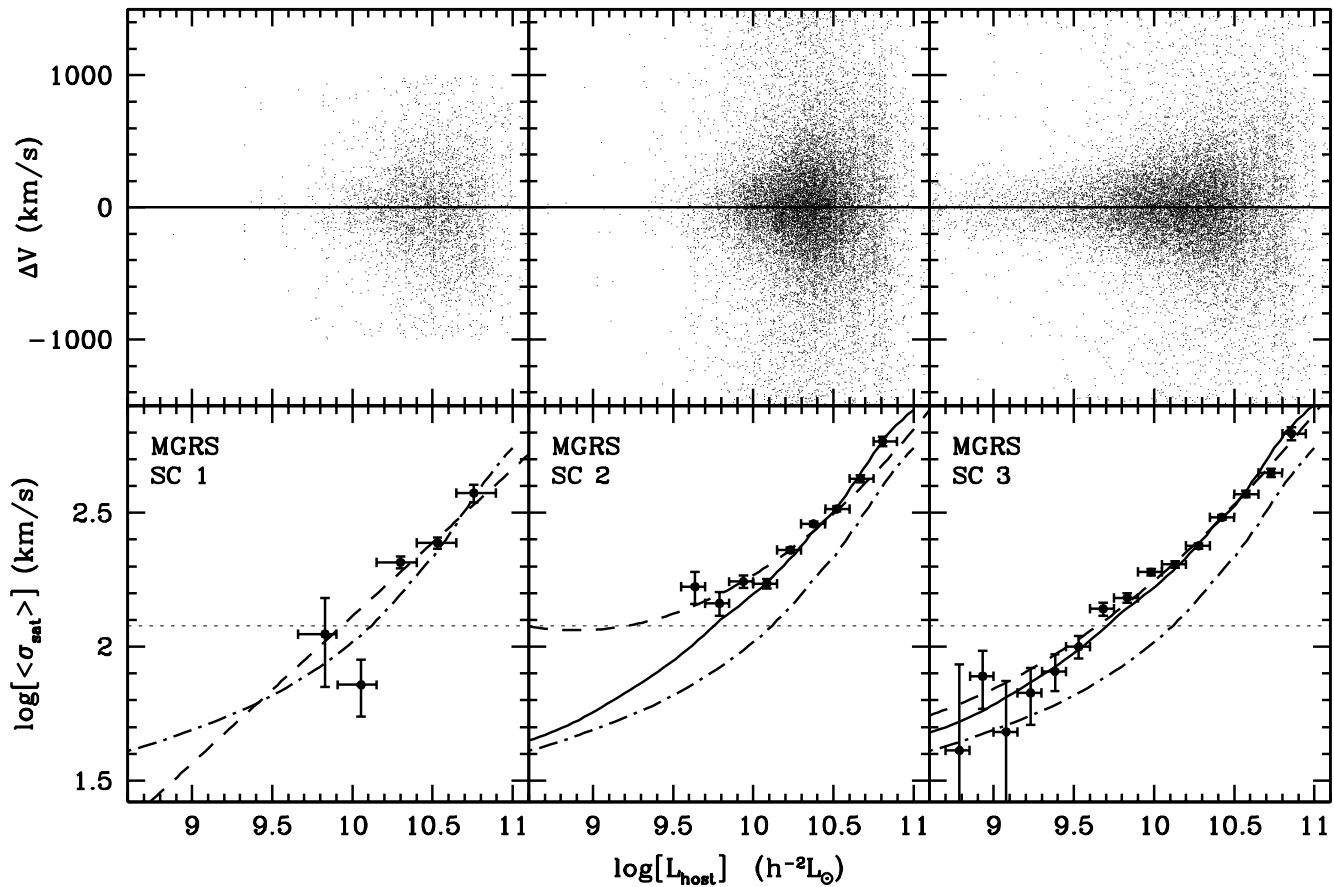,width=\hdsize}}
\caption{Kinematics of satellite galaxies in the MGRS. Upper panels
  show scatter plots of $\Delta V$ versus $L_{\rm host}$ for the three
  different host-satellite samples discussed in the text. Lower panels
  plot the  corresponding satellite velocity  dispersions.  Solid dots
  with  errorbars   indicate  the  $\sigma_{\rm   sat}(L_{\rm  host})$
  obtained by fitting  $P(\Delta V)$ with the sum of  a Gaussian and a
  constant term.  Vertical errorbars  are obtained from the covariance
  matrix of  the Levenberg-Marquardt method used for  the fitting, and
  horizontal  errorbars indicate the  range of  host luminosity  used. 
  The  dashed  lines indicate  the  best-fit $\sigma_{\rm  sat}(L_{\rm
    host})$  obtained  using   the  maximum-likelihood  method.   Open
  circles in the  lower left-hand panel, and solid  lines in the lower
  middle and right-hand panels,  indicate the expectation values.  Any
  difference  between the $\sigma_{\rm  sat}(L_{\rm host})$  and these
  expectation  values   is  due  to   interlopers,  satellites  around
  non-central   hosts,  and   shot  noise   (see  text   for  detailed
  discussion). The  dot-dashed line indicates  the host-averaged mean,
  and is shown for  comparison.  The dotted, horizontal line indicates
  $\sigma_{\rm sat}  = 120 \kms$ which corresponds  to the `resolution
  limit' (due to the errors on $\Delta V$).}
\label{fig:sckin}
\end{figure*}

The upper panels of  Fig~\ref{fig:sckin} show scatter plots of $\Delta
V \equiv V_{\rm sat} - V_{\rm  host}$ as a function of $L_{\rm host}$,
which represents the  `raw data' on satellite kinematics  that we seek
to  quantify. We  follow McKay  \etal (2002)  and Brainerd  \& Specian
(2003) and proceed as  follows.  We bin all host-satellite  pairs in a
number of  bins of $L_{\rm  host}$ and fit the  distribution $P(\Delta
V)$ for  each of these bins with  the sum of a  Gaussian (to represent
the true satellites)  plus a constant (to represent  the interlopers). 
The final  estimate of the projected velocity  dispersion of satellite
galaxies, $\sigma_{\rm sat}$, follows  from the velocity dispersion of
the best-fit  Gaussian after correcting for  the error in  $\Delta V$. 
In our attempt  to mimic the 2dFGRS, we added a  Gaussian error of $85
\kms$ (see Colless  \etal 2001) to the velocity of  each galaxy in our
MGRS. The error  on $\Delta V$ is therefore  equal to $\sqrt{2} \times
85  \kms = 120  \kms$, which  we subtract  from $\sigma_{\rm  sat}$ in
quadrature.  The solid  circles with errorbars in the  lower panels of
Fig.~\ref{fig:sckin}   plot  the   $\sigma_{\rm   sat}(L_{\rm  host})$
obtained  for the  three samples  extracted from  our  MGRS.  Vertical
errorbars   are   obtained  from   the   covariance   matrix  of   the
Levenberg-Marquardt method  used to fit  the Gaussian-plus-constant to
$P(\Delta  V)$,  and  horizontal   errorbars  indicate  the  range  of
luminosities of the host galaxies in each bin.  The dotted, horizontal
line indicates the `resolution limit' of $120 \kms$.

The  constant  term  in   the  fitting  function  has  typically  been
interpreted  as  representing  the  contribution due  to  interlopers
(McKay \etal 2002;  Brainerd \& Specian 2003). However,  we have shown
above  that $P_{\rm int}(\Delta  V)$ differs  strongly from  a uniform
distribution.   The  $f_{\rm  int}(L_{\rm  host})$ that  follows  from
integrating the constant term is shown  as a dashed line in the panels
in  the  middle  column  of  Fig.~\ref{fig:sc}.   As  expected,  these
interloper fractions  are systematically too low compared  to the true
interloper fractions (open circles). Unfortunately, with real data the
detailed $P_{\rm  int}(\Delta V)$ is  unknown, making it  difficult to
properly correct for the  interlopers. We therefore devised a strategy
that  aims at tuning  the selection  criteria to  limit the  number of
interlopers  to acceptable  levels (cf.   SC  3). In  order to  remove
interlopers  with large  $\vert  \Delta  V \vert$  we  still take  the
constant term  into account in the  fitting function.  As  long as the
fraction   of  interlopers  is   sufficiently  small,   the  remaining
interlopers should not strongly effect the kinematics.

In addition to discrete measurements of $\sigma_{\rm sat}$ for several
independent bins  in $L_{\rm  host}$, we also  use a method  that fits
$\sigma_{\rm   sat}(L_{\rm  host})$   to   all  host-satellite   pairs
simultaneously.   We parameterize  the  relation between  $\sigma_{\rm
  sat}$ and $L_{\rm host}$ with a quadratic form in the logarithm:
\begin{equation}
\label{siglum}
{\rm log} \sigma_{\rm sat} =  {\rm log} \sigma_{10} + a_1
{\rm log} L_{10}  + a_2 ({\rm log} L_{10})^2 
\end{equation}
Here $L_{10} = L_{\rm host} / 10^{10} h^{-2} \Lsun$ and $\sigma_{10} =
\sigma_{\rm sat}(L_{10})$.   Let $f_{\rm int}$ denote  the fraction of
interlopers, and  assume that $f_{\rm int}$ is  independent of $L_{\rm
  host}$ and/or  $\Delta V$.  Then,  the probability that  a satellite
around a host with luminosity $L_{\rm host}$ has a velocity difference
with respect to the host of $\Delta V \kms$ is given by
\begin{equation}
\label{prob}
P(\Delta V)  = {f_{\rm  int} \over  2 (\Delta V)_s} +  
{(1 - f_{\rm int}) \over \varpi} \; {\rm exp}\left[
-{\Delta V^2 \over 2 \sigma^2_{\rm eff}}\right]
\end{equation}
Here $\sigma_{\rm  eff} = \sqrt{ \sigma_{\rm sat}^2  + 120^2}$ defines
the  {\it effective}  dispersion that  takes account  of  the velocity
errors, and
\begin{equation}
\label{zet}
\varpi = \sqrt{2 \pi} \, \sigma_{\rm eff} \, 
{\rm erf}\left[ (\Delta V)_s \over 
\sqrt{2} \, \sigma_{\rm eff} \right]
\end{equation}
so that~(\ref{prob})  is properly normalized  to unity over  the valid
range  $\vert \Delta  V \vert  \leq  (\Delta V)_s$.   We use  Powell's
directional     set     method      to     find     the     parameters
$(\sigma_{10},a_1,a_2,f_{\rm  int})$   that  maximize  the  likelihood
${\cal L} \equiv \sum_{i} {\rm ln}[P(\Delta V_i)]$ where the summation
is over all  satellites. It is this maximum  likelihood method that we
use to parameterize the satellite kinematics in our adaptive selection
criterion   (SC    3)   introduced   in    Section~\ref{sec:select}.   
Columns~(8)--(11)  of Table~2  list  the best-fit  values for  $f_{\rm
  int}$  (to be  compared to  the true  interloper fraction  listed in
column~7), $\sigma_{10}$,  $a_1$ and  $a_2$ obtained from  fitting the
host-satellite pairs extracted from the  MGRS. The dashed lines in the
lower  panels  of   Fig.~\ref{fig:sckin}  indicate  the  corresponding
best-fit $\sigma_{\rm sat}(L_{\rm host})$.
\begin{figure*}
\centerline{\psfig{figure=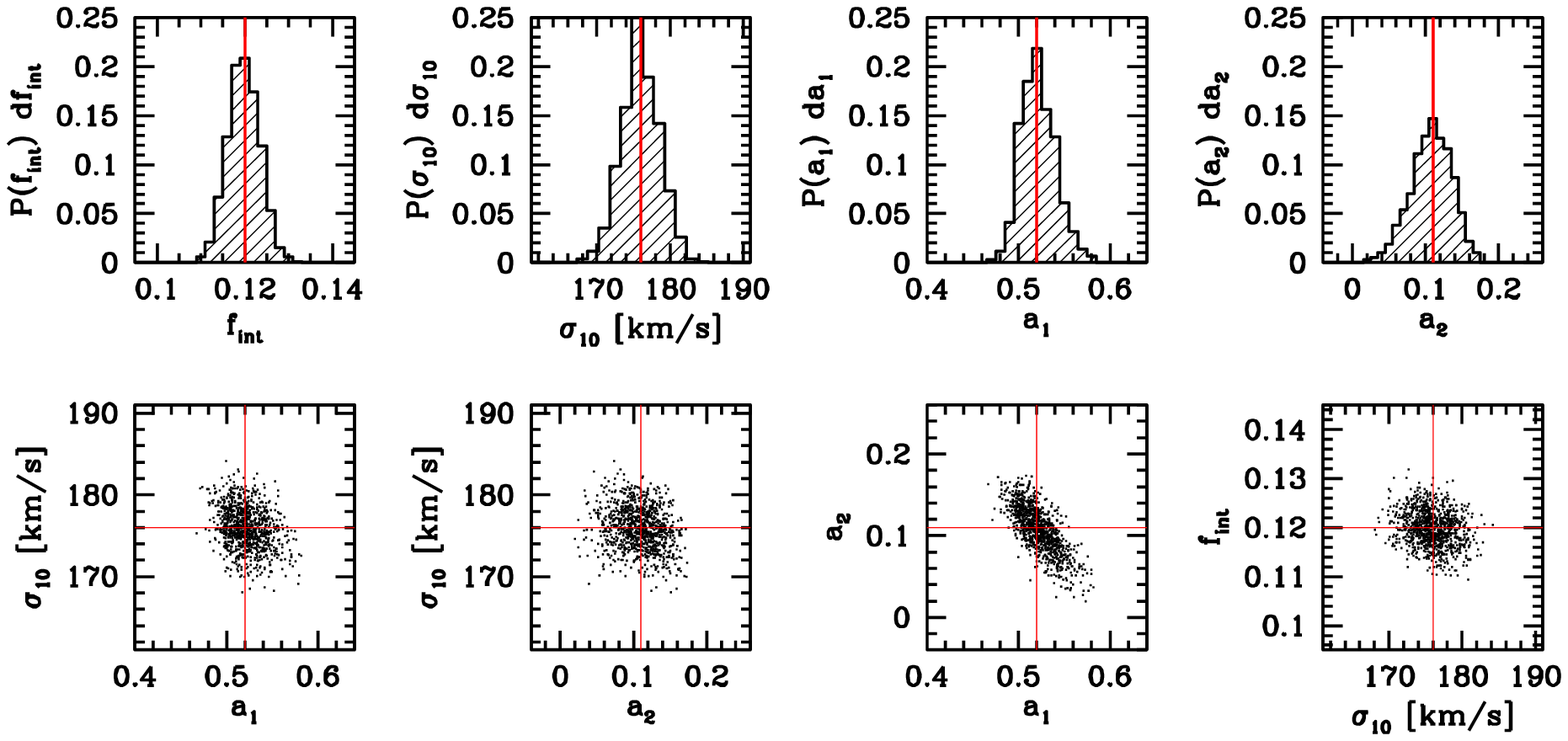,width=\hdsize}}
\caption{The upper panels plot  distributions of $f_{\rm int}$,
  $\sigma_{10}$, $a_1$,  and $a_2$  (as indicated) obtained  from 1000
  Monte-Carlo simulations. These are  used to estimate errors on these
  parameters  obtained from  fitting the  distribution of  $(\Delta V,
  L_{\rm  host})$  in the  MGRS  (see  text).   Thick, vertical  lines
  indicate the  input values.   The lower panels  show two-dimensional
  scatter plots  of these  parameters.  Horizontal and  vertical lines
  again indicate the input values.}
\label{fig:moncar}
\end{figure*}

With SC  1 the  satellite kinematics can  only be  measured accurately
over about  a factor 5 in  luminosity (cf. Brainerd \&  Specian 2003). 
Therefore,  when fitting  $\sigma_{\rm sat}(L_{\rm  host})$  using the
maximum  likelihood method  we keep  $a_2$  fixed at  zero, such  that
eq.~(\ref{siglum}) reduces to a simple  power-law. As can be seen, the
maximum likelihood method and the discrete Gaussian-plus-constant fits
yield  satellite  velocity dispersions  in  good  agreement with  each
other. Because $f_h$ and $f_s$ are not equal to unity with SC 1, it is
difficult to  compute expectation values  for $\sigma_{\rm sat}(L_{\rm
  host})$  based  on  the CLF.   In  order  to  assess the  impact  of
interlopers and  non-central hosts  we therefore compute  the velocity
dispersion of the true satellite galaxies directly from the MGRS: open
circles in the upper-left  panel of Fig.~\ref{fig:sckin} correspond to
$\sqrt{{1 \over  N}\sum_{i=1}^{N} (\Delta  V_i)^2 - 120^2}$  where the
summation  is  over all  true  satellites  (excluding interlopers  and
satellites  around  non-central  hosts).   The  best-fit  $\sigma_{\rm
  sat}(L)$, both  from the maximum  likelihood method as well  as from
the discrete  Gaussian-plus-constant fits, are in  good agreement with
these  true  values, indicating  that  the  incomplete correction  for
interlopers does not significantly influence the satellite kinematics.

In  the  case  of SC  2,  the  luminosity  range over  which  accurate
measurements of $\sigma_{\rm sat}(L)$ can be obtained has increased to
almost one and a half  orders of magnitude.  The $\sigma_{\rm sat}(L)$
is in reasonable agreement  with the expectation values computed using
eq.~(\ref{expsigflux}) with  $R_p =  r_{\rm vir}$ (thick  solid line),
except  for the  lowest luminosity  bins.  This  is due  to  the large
fraction  of (excess)  interlopers and  to the  presence  of satellite
galaxies around non-central hosts.   Especially the latter can cause a
significant   overestimate  of   the  true   $\sigma_{\rm   sat}(L)$.  
Nevertheless,  despite  an interloper-fraction  of  39  percent, SC  2
allows  one  to  recover   the  expected  $\sigma_{\rm  sat}(L)$  with
reasonable accuracy.

The results  for SC 3  are even more  promising. Because of  the large
number of faint host galaxies, $\sigma_{\rm sat}(L_{\rm host})$ can be
measured  over two  and a  half orders  of magnitude  down to  $\sim 4
\times  10^8 h^{-2} \Lsun$.   The $\sigma_{\rm  sat}$ obtained  are in
good agreement with the expectation values (solid line, computed using
eq.~(\ref{expsigflux}) with $R_p  = 0.375 r_{\rm vir}$) ,  even in the
regime where $\sigma_{\rm sat} < 120 \kms$.  The dashed line indicates
the  best-fit  $\sigma_{\rm  sat}(L_{\rm  host})$  obtained  from  the
maximum-likelihood method,  the corresponding parameters  of which are
listed in Table~2.

Fitting  eq.~(\ref{siglum})  directly  to  the  expected  $\sigma_{\rm
  sat}(L_{\rm  host})$ (solid line  in the  lower right-hand  panel of
Fig.~\ref{fig:sckin}) yields  as best-fit parameters: $\sigma_{10}=178
\kms$, $a_1 = 0.58$, and $a_2=0.10$. Thus, the expectation is somewhat
steeper than the actual best-fit relation (for which $a_1 = 0.52$). In
order to  address the  significance of this  difference we  proceed as
follows: we construct Monte-Carlo  samples based on $\sigma_{10} = 176
\kms$,  $a_1=0.52$,   $a_2  =  0.11$,   and  $f_{\rm  int}   =  0.12$,
corresponding to the best-fit values  obtained for the MGRS.  For each
of the 16841 satellite galaxies in the MGRS we randomly draw a $\Delta
V$  from equation~(\ref{prob}), using  $f_{\rm int}$  and $\sigma_{\rm
  sat}(L_{\rm host})$ given by  eq.~(\ref{siglum}). To this $\Delta V$
we add  a Gaussian  deviate with standard  deviation of $120  \kms$ to
mimic the velocity errors.  This yields a sample of $(\Delta V, L_{\rm
  host}$) with  the same distributions  of $L_{\rm host}$  and $N_{\rm
  sat}$ as for the MGRS.   Next we apply the maximum likelihood method
and find  the best-fit values of  $(\sigma_{10},a_1,a_2,f_{\rm int})$. 
The  distributions of  these best-fit  values, obtained  from  1000 of
these  Monte-Carlo  samples,   are  shown  in  Fig.~\ref{fig:moncar}.  
Clearly, the  maximum-likelihood method accurately  recovers the input
values of $\sigma_{10}$, $a_1$, $a_2$, and $f_{\rm int}$ (indicated by
vertical and horizontal lines) with  $1 \sigma$ errorbars of $3 \kms$,
$0.02$, $0.04$ and $0.005$, respectively (and with the errors on $a_1$
and  $a_2$  somewhat  correlated).   Given these  random  errors,  the
difference  between   the  dashed   and  solid  lines   is  marginally
significant,  reflecting  the   effect  of  interlopers  and  (mainly)
satellites around  non-central hosts. Nevertheless,  the difference is
sufficiently small, that we conclude that it is possible to use large,
flux  limited redshift  surveys to  obtain accurate  estimates  of the
velocity dispersion of  satellite galaxies in haloes that  span a wide
range  of masses. However,  keep in  mind that  these are  biased with
respect to the host-averaged means.

\section{Results for the 2\lowercase{d}FGRS}
\label{sec:2dFres}

We now focus on real data.  We use the final, public data release from
the  2dFGRS, restricting  ourselves  only to  galaxies with  redshifts
$0.01 \leq  z \leq 0.15$  in the North  Galactic Pole (NGP)  and South
Galactic Pole  (SGP) subsamples with  a redshift quality  parameter $q
\geq  3$.  This  leaves  a grand  total  of $146735$  galaxies with  a
typical  rms  redshift  error  of  $85 \kms$  (Colless  \etal  2001).  
Absolute magnitudes for galaxies in  the 2dFGRS are computed using the
K-corrections of Madgwick \etal (2002).
\begin{figure}
\centerline{\psfig{figure=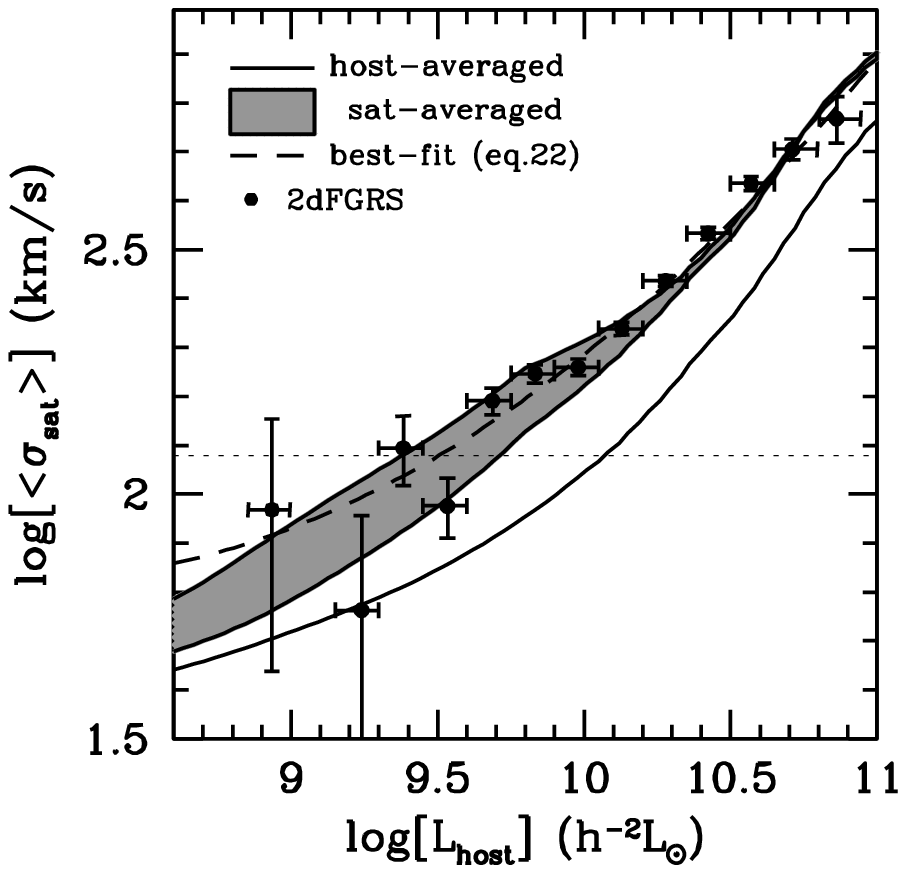,width=\hssize}}
\caption{Solid dots with errorbars  indicate the best-fit 
  $\sigma_{\rm sat}(L_{\rm host})$ obtained  from the 2dFGRS data. The
  dashed  line indicates  the corresponding  best-fit function  of the
  form~(\ref{siglum})  obtained using  the  maximum likelihood  method
  described  in the  text.  The  gray area  indicates  the expectation
  values obtained  using the CLF,  where the upper and  lower boundary
  correspond  to  `random'   and  `constrained'  luminosity  sampling,
  respectively.    The   solid   line  indicates   the   corresponding
  host-averaged mean $\sigma_{\rm sat}(L_{\rm host})$ and is shown for
  comparison.}
\label{fig:comp}
\end{figure}

\subsection{Satellite Kinematics}
\label{sec:dyn}

Applying our adaptive selection criterion  (SC 3) to the 2dFGRS yields
8132 host  galaxies and 12569  satellite galaxies. The first  thing to
notice is  that these  numbers are  quite a bit  smaller than  for the
MGRS.   In order  to  check  whether this  is  consistent with  cosmic
variance,  we  constructed  four  independent MGRSs,  using  different
$L_{100}$ and  $L_{300}$ simulation  boxes. Applying SC  3 to  each of
these MGRSs yields  $N_{\rm host} = 10600 \pm 300$  and $N_{\rm sat} =
16300 \pm 550$. Clearly, the deficit of host and satellite galaxies in
the 2dFGRS compared  to our MGRS is very  significant.  We address the
abundances of  host and  satellite galaxies in  much more detail  in a
forthcoming  paper (van  den  Bosch \etal,  in  preparation). For  the
moment,  however,  we  ignore   this  discrepancy  and  focus  on  the
kinematics only.

Results are  shown in  Fig.~\ref{fig:comp}. Solid dots  with errorbars
indicate the  $\sigma_{\rm sat}(L_{\rm  host})$ obtained by  fitting a
Gaussian-plus-constant.  The thick  dashed line indicates the best-fit
$\sigma_{\rm sat}(L_{\rm host})$ obtained using the maximum likelihood
method  (with the best-fit  parameters listed  in Table~2).   The gray
area indicates the expectation values  obtained from our CLF where the
upper and  lower boundaries  correspond to `random'  and `constrained'
luminosity sampling, respectively.   Clearly, the satellite kinematics
obtained  from  the  2dFGRS  are  in excellent  agreement  with  these
predictions.  For comparison, the solid line indicates the expectation
value    obtained   using    host-averaging    (see   discussion    in
Section~\ref{sec:analest}).  As  already discussed, the  flux limit of
the 2dFGRS  and the satellite-averaging results in  a significant bias
of  the measured  $\sigma_{\rm  sat}(L_{\rm host})$  compared to  this
host-averaged mean.

The expectation  values (gray area)  are computed assuming  no spatial
bias between satellite galaxies and the dark matter mass distribution.
As  we have  shown in  Section~\ref{sec:bias}, the  effect  of spatial
bias,  and the  resulting  velocity  bias, is  small  compared to  the
uncertainties resulting from the luminosity sampling.  In other words,
deviations  of the  true $n_{\rm  sat}(r)$ from  the  NFW distribution
assumed here  will not significantly  affect our main  conclusion that
the satellite kinematics of the 2dFGRS are in excellent agreement with
predictions based on our CLF. 

\section{Summary}
\label{sec:summ}

Previous attempts to measure the kinematics of satellite galaxies have
mainly  focussed on  isolated  spiral galaxies.   Using detailed  mock
galaxy  redshift surveys  we  investigated to  what  extent a  similar
analysis can be  extended to include a much  wider variety of systems,
from isolated galaxies to massive  groups and clusters.  Our method is
based  on the  assumption that  satellite galaxies  are  an isotropic,
steady-state  tracer  population orbiting  within  spherical NFW  dark
matter haloes, and  that the brightest galaxy in  each halo resides at
rest at the  halo centre.  These assumptions are,  at least partially,
supported  by both  observations of  cluster galaxies  (i.e.,  van der
Marel  \etal  2000)  and  by  numerical  simulations  of  dark  matter
subhaloes (see  Diemand \etal 2004 and  references therein).  Although
there  are most  definitely  systems in  which  one or  more of  these
assumptions break  down, we hypothesized that  with sufficiently large
samples   of   host-satellite   pairs  the   occasional   `perturbed',
non-relaxed, system  will not significantly influence  the results. In
addition, we have  shown that the assumption of  orbital isotropy only
influences the  velocity dispersion of  satellite galaxies at  the few
percent level.

We used  the conditional luminosity  function (CLF) formalism  to make
predictions of the observed  velocity dispersion of satellite galaxies
around  host galaxies  of different  luminosity.  We  showed  that the
satellite  weighting,  which   occurs  naturally  when  stacking  many
host-satellite  pairs to increase  signal-to-noise, introduces  a bias
towards higher $\sigma_{\rm sat}(L_{\rm  host})$ compared to the true,
host-averaged  mean.   A  further  bias,  in the  same  direction,  is
introduced  when   using  flux-limited,  rather   than  volume-limited
surveys.   Finally, we demonstrated  that most  of the  uncertainty in
interpreting the measured $\sigma_{\rm sat}(L_{\rm host})$ owes to the
unknown second moment of the conditional probability distribution $P(M
\vert L_{\rm  host})$: typically a larger second  moment yields higher
expectation values.  The  CLF formalism is the ideal  tool to properly
take all these various biases into account.
 
An important, additional problem  with the interpretation of satellite
kinematics  is how  to deal  with the  presence of  interlopers (i.e.,
galaxies  selected   as  satellites  but  which   are  not  physically
associated with  the halo  of the host  galaxy) and  non-central hosts
(i.e., galaxies that are selected  as host galaxies, but which are not
the central,  brightest galaxy in  their own halo). The  abundances of
interlopers  and non-central  hosts depend  strongly on  the selection
criteria.   Using MGRSs,  constructed  from the  CLF, we  investigated
different selection criteria for  hosts and satellites.  The first one
is identical to  the criteria used by McKay  \etal (2002) and Brainerd
\& Specian  (2003) in  their analyses of  satellite kinematics  in the
SDSS and  2dFGRS, respectively.   Applied to our  MGRS it  yields only
3876 satellites, which only allows an analysis of satellite kinematics
around host galaxies that span  a factor 5 in luminosity. Although the
fraction of  non-central hosts is,  with one percent,  negligible, the
fraction  of  interlopers  is   27  percent.   More  importantly,  the
interloper    fraction     increases    strongly    with    decreasing
host-luminosity,  reaching values  as high  as 60  percent  around the
faintest host galaxies. Contrary to  what has been assumed in the past
(e.g.,  McKay \etal  2002;  Brainerd \&  Specian  2003), the  velocity
distribution of interlopers is strongly peaked towards $\Delta V = 0$,
similar to the distribution of  true satellite galaxies.  This makes a
detailed correction for interlopers  extremely difficult. In fact, the
method used thus  far, based on the assumption  of a uniform $P(\Delta
V)$, typically underpredicts the true interloper fraction by $\sim 50$
percent.

We therefore  devised an iterative, adaptive  selection criterion that
yields large numbers  of hosts and satellites with  as few interlopers
and non-central hosts as possible. Applying these criteria to our MGRS
yields  16750  satellites  around   10483  hosts.   In  addition,  the
interloper fraction is only 14 percent and, more importantly, does not
vary significantly  with host luminosity.  Because of  the much larger
numbers involved,  and the reduced fraction  of interlopers, satellite
kinematics can be analyzed over two  and a half orders of magnitude in
$L_{\rm  host}$.  The resulting  $\sigma_{\rm  sat}(L_{\rm host})$  is
found to be in good  agreement with the expectation values, indicating
that  the  interlopers  and  non-central hosts  do  not  significantly
distort the measurements.  

Finally, we applied our adaptive selection criteria to the 2dFGRS. The
resulting  satellite  kinematics   are  in  excellent  agreement  with
predictions  based  on the  CLF  formalism,  once  the various  biases
discussed above  are taken into  account.  We therefore  conclude that
the  observed  kinematics  of  satellite  galaxies  provide  virtually
independent,  dynamical  confirmation  of  the  average  mass-to-light
ratios inferred by Yang \etal  (2003a) and van den Bosch \etal (2003a)
from  a  purely  statistical   method  based  on  the  abundances  and
clustering properties of galaxies in the 2dFGRS.


\section*{Acknowledgements}

We are grateful  to Yipeng Jing for providing us  the set of numerical
simulations  used for  the construction  of our  mock  galaxy redshift
surveys,  and  to  Darren  Croton,  Ben  Moore,  Juerg  Diemand,  Anna
Pasquali, Simon  White, and the anonymous referee  for useful comments
and discussions.



\appendix

\section[]{The conditional luminosity function}
\label{sec:AppA}

The  construction  of  our  mock  galaxy  redshift  surveys  uses  the
conditional luminosity function (CLF) to indicate how many galaxies of
given luminosity  occupy a halo of  given mass. The  CLF formalism was
introduced by Yang \etal (2003a)  and van den Bosch \etal (2003a), and
we refer the reader to these  papers for more details. Here we briefly
summarize  the main  ingredients and  we present  the parameterization
used.

The CLF is parameterized by a Schechter function:
\begin{equation}
\label{phiLM}
\Phi(L  \vert  M)  {\rm  d}L  = {\tilde{\Phi}^{*}  \over  \wLstar}  \,
\left({L \over  \wLstar}\right)^{\walpha} \, \,  {\rm exp}(-L/\wLstar)
\, {\rm d}L,
\end{equation}
where   $\wLstar   =   \wLstar(M)$,   $\walpha   =   \walpha(M)$   and
$\tilde{\Phi}^{*}  = \tilde{\Phi}^{*}(M)$  are all  functions  of halo
mass $M$. We write the average, total mass-to-light ratio of a halo of
mass $M$ as
\begin{equation}
\label{MtoLmodel}
\left\langle {M \over L} \right\rangle(M) = {1 \over 2} \,
\left({M \over L}\right)_0 \left[ \left({M \over M_1}\right)^{-\gamma_1} +
\left({M \over M_1}\right)^{\gamma_2}\right],
\end{equation}
This parameterization has four  free parameters: a characteristic mass
$M_1$, for  which the mass-to-light  ratio is equal to  $(M/L)_0$, and
two slopes,  $\gamma_1$ and $\gamma_2$,  that specify the  behavior of
$\langle M/L  \rangle$ at  the low and  high mass ends,  respectively. 
Motivated  by observations  (Bahcall,  Lubin \&  Norman 1995;  Bahcall
\etal 2000; Sanderson \& Ponman  2003; Eke \etal 2004), which indicate
a  flattening of  $\langle  M/L  \rangle(M)$ on  the  scale of  galaxy
clusters, we set $\langle M/L  \rangle(M) = (M/L)_{\rm cl}$ for haloes
with $M \geq 10^{14}  h^{-1} \Msun$.  With $(M/L)_{\rm cl}$ specified,
the value for $\gamma_2$ derives from requiring continuity in $\langle
M/L \rangle(M)$ across $M = 10^{14} h^{-1} \Msun$.

A  similar parameterization is used  for the characteristic luminosity
$\wLstar(M)$:
\begin{equation}
\label{LstarM}
{M \over \wLstar(M)} = {1 \over 2} \, \left({M \over L}\right)_0 \,
f(\walpha) \, \left[ \left({M \over M_1}\right)^{-\gamma_1} +
\left({M \over M_2}\right)^{\gamma_3}\right],
\end{equation}
with
\begin{equation}
\label{falpha}
f(\walpha) = {\Gamma(\walpha+2) \over \Gamma(\walpha+1,1)}.
\end{equation}
Here  $\Gamma(x)$   is  the  Gamma  function   and  $\Gamma(a,x)$  the
incomplete Gamma  function.  This parameterization  has two additional
free  parameters: a characteristic  mass $M_2$  and a  power-law slope
$\gamma_3$.   For $\walpha(M)$ we  adopt a  simple linear  function of
$\log(M)$,
\begin{equation}
\label{alphaM}
\walpha(M) = \alpha_{15} + \eta \, \log(M_{15}),
\end{equation}
with $M_{15}$ the halo mass in units of $10^{15} \msunh$, $\alpha_{15}
= \walpha(M_{15}=1)$, and $\eta$ describes the change of the faint-end
slope $\walpha$ with halo mass.   Finally, we introduce the mass scale
$M_{\rm min}$ below which we set the CLF to zero; i.e., we assume that
no  stars form inside  haloes with  $M <  M_{\rm min}$.   Motivated by
reionization  considerations (see  Yang  \etal 2003a  for details)  we
adopt $M_{\rm min} = 10^{9} h^{-1} \Msun$ throughout.

In  this paper  we use  a CLF  with the  following parameters:  $M_1 =
10^{11.12} h^{-1} \Msun$, $M_2=10^{11.71} h^{-1} \Msun$, $(M/L)_0=85 h
\MLsun$,     $\gamma_1=1.55$,     $\gamma_2=0.46$,    $\gamma_3=0.69$,
$\eta=-0.29$  and $\alpha_{15}=-0.99$.  This  model is  different from
those listed  in van den  Bosch \etal (2003a)  as it yields  a higher,
average mass-to-light  ratio on  the scale of  clusters.  As  shown in
Yang  \etal (2003b),  this is  in better  agreement with  the observed
pairwise peculiar  velocity dispersions  of 2dFGRS galaxies  (see also
van  den Bosch,  Mo  \& Yang  2003b).   However, none  of the  results
presented in  this paper are  sensitive to our  choice of the  CLF. We
verified that  MGRSs based  on either  of the CLFs  listed in  van den
Bosch  \etal  (2003a)  yield  virtually  identical  results  to  those
presented here.

\section[]{Probability distribution of luminosity of  
  brightest galaxy in a dark matter halo.}
\label{sec:AppB}

Define $L_c$ as the luminosity of the {\it brightest} galaxy in a halo
of mass  $M$. It  is convenient to  write the  conditional probability
distribution $P(L_c \vert  M) {\rm d}L_c$ in terms  of the conditional
luminosity function $\Phi(L_c \vert M)  {\rm d}L_c$ and a new function
$f(L_c,M)$ which depends on how galaxy luminosities are drawn from the
CLF:
\begin{equation}
\label{probLbright}
P(L_c \vert M) {\rm d}L_c = \Phi(L_c \vert M) \, f(L_c,M) \, {\rm d}L_c
\end{equation}
In the case  of `constrained' drawing, $L_c$ has  an expectation value
given by eq.~(\ref{Lcentral}), and it is straightforward to show that
\begin{equation}
\label{casef}
f(L_c,M) = \left\{ \begin{array}{ll}
1 & \mbox{if $L_c \geq L_1(M)$} \\
0 & \mbox{if $L_c < L_1(M)$}
\end{array} \right.
\end{equation}
with $L_1(M)$ as defined by eq.~(\ref{Lj}). Note that in the MGRS used
in this paper  the luminosity of the brightest galaxy  in each halo is
always  drawn  constrained.   Therefore,  when  computing  expectation
values for $\sigma_{\rm sat}(L_{\rm host})$ in the MGRS, we use $P(L_c
\vert M)$ with $f(L_c,M)$ given by eq.~(\ref{casef}).

In the  case of `random' drawing,  the situation is  more complicated. 
The  probability that  a galaxy  drawn at  random from  the CLF  has a
luminosity less than $L_c$ is given by
\begin{equation}
\label{cumprobone}
P(< L_c \vert M) = 1 - {1 \over \langle N \rangle_M}
  \int_{L_c}^{\infty} \Phi(L \vert M) {\rm d}L
\end{equation}
with $\langle  N \rangle_M$ the mean  number of galaxies in  a halo of
mass $M$ given by eq.~(\ref{meanN}).  In a halo with $N$ galaxies, the
probability that the brightest galaxy has $L < L_c$ is simply $[P(<L_c
\vert  M)]^N$.   Differentiating  with  respect to  $L_c$  yields  the
probability $P_N(L_c \vert M) {\rm  d}L_c$ that after $N$ drawings the
brightest galaxy  has a luminosity in  the range $L_c  \pm L_c/2$. The
full  probability $P(L_c  \vert M)  {\rm d}L_c$  follows  from summing
$P_N(L_c  \vert M)  {\rm d}L_c$  over  $N$, properly  weighted by  the
probability  $P(N \vert  M)$  that a  halo  of mass  $M$ contains  $N$
galaxies. This yields
\begin{equation}
\label{cumprobN}
f(L_c,M) = {1 \over \langle N \rangle_M} \sum_{N=1}^{\infty} 
N \, P(N \vert M) \,  \left[ P(<L_c \vert  M)\right]^{N-1}
\end{equation}
If $\langle N \rangle_M \leq 1$ then
\begin{equation}
\label{pnm}
P(N \vert M) = \left\{ \begin{array}{ll}
1 - \langle N \rangle_M & \mbox{if $N=0$} \\
\langle N \rangle_M     & \mbox{if $N=1$}
\end{array} \right.
\end{equation}
(see Section~\ref{sec:hon}),  so that $f(L_c,M) = 1$.   For $\langle N
\rangle_M > 1$ we have that
\begin{equation}
\label{pnmpoi}
P(N \vert M) =  {(\langle N \rangle - 1)^{N-1} \over (N-1)!} \, 
{\rm exp}(1 - \langle N \rangle_M) 
\end{equation}
(see        Section~\ref{sec:hon}).        Substituting~(\ref{pnmpoi})
in~(\ref{cumprobN}) and using that
\begin{equation}
\label{gr}
\sum_{k=0}^{\infty} {x^k \over k!} (k+1) = (1+x) \, {\rm exp}(x)
\end{equation}
(see eq.~[1.212] in Gradshteyn \& Ryzhik, 1980) yields 
\begin{equation}
\label{fran}
f(L_c,M) = \left( 1 - {\zeta \over \langle N \rangle_M} \right) \,
{\rm exp}(-\zeta)
\end{equation}
with 
\begin{equation}
\label{zeta}
\zeta = {\langle N \rangle_M - 1 \over \langle N \rangle_M} \,
\int_{L_c}^{\infty} \Phi(L \vert M) {\rm d}L
\end{equation}

\label{lastpage}


\begin{thebibliography}{}

\bibitem[]{Bah95}
Bahcall N.A., Lubin L., Dorman V., 1995, \apj , 447, L81

\bibitem[]{Bah00}
Bahcall N.A., Cen R., Dav\'e R., Ostriker J.P., Yu Q., 2000, \apj ,
541, 1

\bibitem[]{Ben00}
Benson A.J., Cole S., Frenk C.S., Baugh C.M., Lacey C.G., 2000,
\mnras , 311, 793

\bibitem[]{Ber02}
Berlind A.A., Weinberg D.H., 2002, \apj , 575, 587

\bibitem[]{BWB03}
Berlind A.A., et. al., 2003, \apj , 593, 1

\bibitem[]{Bin87}
Binney J.J., Tremaine S.D., 1987, Galactic Dynamics (Princeton:
Princeton Univ. Press)

\bibitem[]{Bra03}
Brainerd T.G., Specian M.A., 2003, \apj , 593, L7

\bibitem[]{Car96}
Carlberg R.G., Yee H.K.C., Ellingson E., Abraham R., Gravel P., Morris
S., Pritchet C.J., 1996, \apj , 462, 32

\bibitem[]{Car97a}
Carlberg R.G., et al., 1997a, \apj , 485, L13

\bibitem[]{Car97b}
Carlberg R.G., Yee H.K.C., Ellingson E., 1997b, \apj , 478, 462

\bibitem[]{Col99}
Colin P., Klypin A., Kravtsov A.V., Khokhlov M., 1999, \apj , 532, 32

\bibitem[]{Col01}
Colless M., The 2dFGRS team, 2001, \mnras , 328, 1039

\bibitem[]{Dav85}
Davis M., Efstathiou G., Frenk C.S., White S.D.M., 1985, \apj , 292, 371

\bibitem[]{DeL04}
De Lucia G., Kauffmann G., Springel V., White S.D.M., Lanzoni B., 
Stoehr F., Tormen G., Yoshida N., 2004, \mnras, 348, 333

\bibitem[]{Die04}
Diemand J., Moore B., Stadel J., 2004, preprint (astro-ph/0402160)

\bibitem[]{Eke01}
Eke V.R., Navarro J.F., Steinmetz M., 2001, \apj , 554, 114

\bibitem[]{Eke04}
Eke V.R., The 2dFGRS team, 2004, preprint (astro-ph/0402566)  

\bibitem[]{Eri89}
Erickson L.K., Gottesman S.T., Hunter J.H., 1987, \nat , 325 779

\bibitem[]{EvW00}
Evans N.W., Wilkinson M.I., 2000, \mnras , 316, 929

\bibitem[]{Eva00}
Evans N.W., Wilkinson M.I., Guhathakurta P., Grebel E.K., Vogt S.S.,
2000, \apj , 540, L9

\bibitem[]{Ghi98}
Ghigna S., Moore B., Governato F., Lake G., Quinn T., Stadel J., 1998,
\mnras , 300, 146

\bibitem[]{GR}
Gradshteyn I.S., Ryzhik I.M., 1980, Table of Integrals, Series, and
Products. Academic Press, New York

\bibitem[]{Ji02}
Jing Y.P., 2002, \mnras , 335, L89

\bibitem[]{JiS02}
Jing Y.P., Suto, Y., 2002, \apj , 574, 538

\bibitem[]{Kly99}
Klypin A., Gottl\"ober S., Kravtsov A.V., Khokhlov A.M., 1999, \apj ,
516, 530

\bibitem[]{Kra03}
Kravtsov A.V., Berlind A.A., Wechsler R.H., Klypin A.A., Gottl\"ober
S., Allgood B., Primack J.R., 2003, preprint (astro-ph/0308519)

\bibitem[]{Lin95}
Lin D.N.C., Jones B.F., Klemola A.R., 1995, \apj , 439, 652

\bibitem[]{Lin04}
Lin Y.-T., Mohr J.J., Stanford S.A., 2004, preprint (astro-ph/0402308)

\bibitem[]{Lit87}
Little B., Tremaine S., 1987, 320, 493

\bibitem[]{Mad90}
Maddox S.J., Efstathiou G., Sutherland W.J., Loveday L., 1990, \mnras
, 243, 692

\bibitem[]{Mea02} 
Madgwick D.S., The 2dFGRS team, 2002, \mnras , 333, 133

\bibitem[]{McK02}
McKay T.A. et al., 2002, \apj , 571, L85

\bibitem[]{Moo98}
Moore B., Governato G., Quinn T., Stadel J., Lake G., 1998, \apj , 499, L5

\bibitem[]{Nav97} 
Navarro J.F., Frenk C.S., White S.D.M., 1997, \apj , 490, 493

\bibitem[]{Nor02a}
Norberg P., The 2dFGRS team, 2002a, \mnras , 332, 827

\bibitem[]{Nor02b}
Norberg P., The 2dFGRS team, 2002b, \mnras , 336, 907

\bibitem[]{Oka99}
Okamoto T., Habe A., 1999, \apj , 516, 591

\bibitem[]{Pra03}
Prada F., et al., 2003, \apj , 598, 260

\bibitem[]{Rhi04}
Rhines K., Geller M.J., Diaferio A., Kurtz M.J., Jarrett T.H., 2004,
preprint (astro-ph/0402242)

\bibitem[]{San03}
Sanderson A.J.R., Ponman T.J., 2003, \mnras , 345, 1241

\bibitem[]{Sco01} 
Scoccimarro R., Sheth R.K., Hui L., Jain B., 2001, \apj , 546, 20

\bibitem[]{Sel00} 
Seljak U.,2000, \mnras , 318, 203

\bibitem[]{SMT01}
Sheth R.K., Mo H.J., Tormen G., 2001a, \mnras , 323, 1

\bibitem[]{ST02} 
Sheth R.K., Tormen, G., 2002, \mnras , 329, 61

\bibitem[]{Spr01}
Springel V., White S.D.M., Tormen G., Kauffmann G., 2001, \mnras , 328, 726

\bibitem[]{Val04}
Vale A., Ostriker J.P., 2004, preprint (astro-ph/0402500)

\bibitem[]{BYM03a}
van den Bosch F.C., Yang X., Mo H.J., 2003a, \mnras , 340, 771

\bibitem[]{BMY03b}
van den Bosch F.C., Mo H.J., Yang X., 2003b, \mnras , 345, 923

\bibitem[]{vdM00}
van der Marel R.P., Magorrian J., Carlberg R.G., Yee H.K.C., Ellingson
E., 2000, \aj , 119, 2038

\bibitem[]{YMB03}
Yang X., Mo H.J., van den Bosch F.C., 2003a, \mnras , 339, 1057

\bibitem[]{YMJBC}
Yang X., Mo H.J., Jing Y.P., van den Bosch F.C., Chu Y., 2003b, 
\mnras , in press (astro-ph/0303524)

\bibitem[]{Yor00} 
York D., et al., 2000, \aj , 120, 1579

\bibitem[]{Zar93}
Zaritsky D., Smith R., Frenk C.S., White S.D.M., 1993, \apj , 405, 464

\bibitem[]{Zar94}
Zaritsky D., White S.D.M., 1994, \apj , 435, 599

\bibitem[]{Zar97}
Zaritsky D., Smith R., Frenk C.S., White S.D.M., 1997, \apj , 478, 39

\bibitem[]{Zwi33}
Zwicky F., 1933, Helv. Phys. Acta, 6, 110

\bibitem[]{Zwi37}
Zwicky F., 1937, \apj , 86, 217

\end{thebibliography}
\end{document}